\documentclass{vgtc}              

\usepackage{amsmath,amsfonts}
\usepackage{amssymb}
 
\title{Predicting Global {HRTF}s From Scanned Head Geometry Using\\ Deep Learning and Compact Representations}

\author{Yuxiang Wang\thanks{e-mail: yuxiang.wang@rochester.edu} %
\and You Zhang\thanks{e-mail: you.zhang@rochester.edu} %
\and Zhiyao Duan\thanks{e-mail: zhiyao.duan@rochester.edu}
\and Mark Bocko\thanks{e-mail: mark.bocko@rochester.edu}}
\affiliation{\scriptsize Department of Electrical and Computer Engineering \\ University of Rochester}

\abstract{%
  In the growing field of virtual auditory display, personalized head-related transfer functions (HRTFs) play a vital role in establishing an accurate sound image for mixed and augmented reality applications. In this work, we propose an HRTF personalization method employing convolutional neural networks (CNN) to predict a subject’s HRTFs for all directions from their scanned head geometry. To ease the training of the CNN models, we propose novel pre-processing methods for both the head scans and HRTF data to achieve compact representations. For the head scan, we use truncated spherical cap harmonic (SCH) coefficients to represent the pinna area, which is important in the acoustic scattering process. For the HRTF data, we use truncated spherical harmonic (SH) coefficients to represent the HRTF magnitudes and onsets. One CNN model is trained to predict the SH coefficients of the HRTF magnitudes from the SCH coefficients of the scanned ear geometry and other anthropometric measurements of the head. The other CNN model is trained to predict SH coefficients of the HRTF onsets from only the anthropometric measurements of the ear, head, and torso. Combining the magnitude and onset predictions, our method is able to predict the complete and global HRTF data. A leave-one-out validation with the log-spectral distortion (LSD) metric is used for objective evaluation. The results show a decent LSD level at both spatial \& temporal dimensions compared to the ground-truth HRTFs and a lower LSD than the boundary element method (BEM) simulation of HRTFs that the database provides. The localization simulation results with an auditory model are also consistent with the objective evaluation metrics, showing the localization responses with our predicted HRTFs are significantly better than with the BEM-calculated ones.
  
}


\renewcommand{\manuscriptnotetxt}{}

\graphicspath{{figs/}{figures/}{pictures/}{images/}{./}} 

\usepackage{tabu}                      
\usepackage{booktabs}                  
\usepackage{lipsum}                    
\usepackage{mwe}                       

\usepackage{mathptmx}                  

\begin{document}


\firstsection{Introduction}

\maketitle

In mixed reality (MR) and augmented reality (AR) applications, binaural rendering aims at providing an immersive environment for the listener. For the successful implementation of AR/VR systems, it is well established that spatial or three-dimensional audio is important for enhancing immersiveness when paired with video~\cite{xie2013head}. Head-related transfer functions (HRTFs) are believed to be a vital tool for spatial auditory display, enabling the listener to efficiently assign direction and distance information to the auditory events~\cite{Blauert1996}. An HRTF set, containing a set of HRTFs with different relative source-listener directions, provides the magnitude, temporal and spectral cues that the auditory system utilizes for spatial perception. For current binaural synthesis engines, most of them are using non-individual generic HRTFs. This is known to cause compromised results such as weak externalization, front-back confusion, and misplaced auditory images. Ideally for each listener, the fully personalized HRTFs are desired. However, due to the fact that there is no easy way to provide one's full HRTF set through measuring or computation~\cite{guezenoc2020hrtf}, effective ways to obtain personalized HRTFs are still under exploration. 

In the past two decades, a great amount of work has been made toward the goal to bring personalized HRTFs, from multiple approaches such as HRTF recommendation/synthesis from key anthropometric features~\cite{Zotkin, Hu2008, Grindlay2007, bilinski2014hrtf}. With the growth of deep learning techniques, in more recent work, HRTF prediction was performed from mesh voxelization \cite{Zhou2021}, auto-encoding ear images and mesh \cite{Miccini2020, Miccini2021, Fantini2021, Yao2022}, notch frequency estimation with ear shape \cite{onofrei20203d}.  

In some previous work~\cite{wang2020global, Xi2021, Zhao2022}, the feasibility of global HRTF prediction was explored in the magnitude domain, using spherical harmonics (SH) as a compact representation method, with anthropometric measurements as inputs. However, the anthropometric measurements, especially for the ear measurements, are subjectively selected parameters, which may produce bias in the representation of the complete head and torso geometry. Given the high complexity of HRTF features, using only a few measurements as the HRTF prediction basis may create overfitting. It is advantageous to derive features possessing distinct mathematical and physical significance, for our method to connect geometry to acoustics dimension sufficiently. Thanks to the progress in optic-based 3D scanning devices, obtaining detailed geometry scans for human head\&torso has become a feasible task, and many of the recent HRTF databases included the scans for each subject~\cite{brinkmann2019hutubs, Engel2023, Armstrong2018, sridhar2017database}. The proposed method for predicting HRTFs using scanned human geometry holds practical potential, enabling AR/VR wearables to generate HRTF predictions based on scans captured by the devices. Still, the scanned head mesh is highly complex data; not all the geometric information on the head is related to the features in HRTFs. In such a case, an analyzing tool is desired to model the geometry within certain regions of the head, since the ear region contributes most to the HRTF spectral cues~\cite{Hebrank1974, Stitt2021}. 

We come to notice spherical cap harmonic analysis (SCHA) as a well-developed technique for regional modeling, which has been extensively studied in the field of geography and computer graphics \cite{Haines1985, Korte2003}. With its adaptation in mesh processing, it has great potential in the efficiency of representing regional geometry, while providing meaningful insight into how the geometry features affect the acoustic features.

In this paper, we present our approach for global HRTF prediction based on an efficient compact representation of both acoustic features in HRTFs and geometry features of the subjects. We use the HUTUBS HRTF database~\cite{brinkmann2019hutubs, Brinkmann2019} as our training data, as it provides a near-uniform source sampling scheme covering the whole sphere, which makes the SH analysis feasible. The scanned meshes of each subject's full head also make the geometry analysis feasible. For the HRTF data, we represent both the magnitude and onset features with spherical-harmonics transform (SHT) coefficients. For the scanned head models, we employ SCHA to efficiently represent the geometry features in the ear regions. A deep learning framework is designed to capture the connections between the acoustic and geometric features extracted in the previous steps. The input of the framework is set as the compact representations of the HRTFs, and the output is set as the compact representations of the meshes. The results show a decent error level at both spatial \& temporal dimensions compared to the ground truth, and have lower error than the acoustic simulation of HRTFs that the database provides. The auditory model experiment results are also consistent with these objective error metrics. 

The main contributions of our work are threefold: (1) We propose a compact global representation for HRTF data using SH expansion for both the magnitude and onset. (2) We propose a novel and compact representation for the subjects' ear geometry using SCH expansion. (3) We design a deep learning framework to predict the SH representation of HRTFs from the SCH representation of ear geometry and other anthropometric features of the upper body, achieving significantly better prediction accuracy compared to the acoustically calculated results. 

The rest of this paper is organized as follows: In Section~\ref{sec: HRTF_pre}, we present our methods for HRTF pre-processing in both frequency and temporal domains. In Section~\ref{sec: compact_represent_mesh}, we demonstrate the process flow we developed for mesh feature extraction. In Section~\ref{sec:Deep_learning_design}, we present the deep learning network design and implementation. The experiments and prediction results are described in Section~\ref{sec:pred_result}, and we conclude our paper in Section~\ref{sec:discussion}. 

\section{HRTF pre-processing}
\label{sec: HRTF_pre}

As HRTFs are high dimensional in both time and frequency domains, direct prediction of them is very challenging. Instead, we pre-process them into more compact representations to make the prediction easier.

\subsection{Compact global representation of HRTF magnitudes}
\label{ssec: compact_glob_hrtf}
In this step, we maintain a similar processing scheme as some previous work \cite{Zotkin2009, ahrens2012hrtf}, processing the HRTFs magnitudes into the real SHT representation. A set of HRTFs can be viewed as functions defined on a spherical surface. The head-related impulse response (HRIR) for each direction is measured between an external sound source and a microphone at the entrance of the ear canal~\cite{xie2013head, algazi2001cipic}. Employing the acoustic reciprocity principle, the same impulse response would be obtained if the source and microphone were to swap their locations. 
Hence, a set of HRTFs can also be viewed as a directivity pattern of a speaker located in the ear canal. To represent such patterns on a spherical surface, spherical harmonics (SH) basis functions are a natural choice, as human listeners are shown to be insensitive to the smoothing of HRTF fine structures~\cite{Kulkarni1998}. 

The Spherical Harmonics (SH) are a set of orthogonal bases for the spherical coordinate system and have been widely adopted in the field of spatial audio. The SHT methods are also used in some recent work in neural representations~\cite{Mller2022, FridovichKeil2022}, to model the directional information. The spherical harmonic basis of the $l$-th order and the $m$-th degree at a certain spatial location is computed as
\begin{equation}
     Y_{l}^{m}(\theta, \varphi)=\sqrt{\frac{(2 l+1)}{4 \pi} \frac{(l-m) !}{(l+m) !}} P_{l}^{m}(\cos \theta) e^{i m \varphi},
\label{eq: SH base eqn}
\end{equation}
where $\theta, \varphi$ are the azimuth and elevation angles in the spherical coordinate system, respectively. $P^{m}_{l}\left( \cos \theta \right)$ is the associated Legendre polynomial. 
 

The SHT process computes coefficients of each SH basis function, and in practice we follow the method in~\cite{Romigh2015}. The SH coefficients are solutions to a system of linear equations~\eqref{eq: linear_eqn1} using $S$ discretized samples at each spatial location $\left\{ \theta _{i},\varphi _{i}\right\} ^{S}$ on the spherical surface:
\begin{equation}
    \boldsymbol{f = Yc}, 
\label{eq: linear_eqn1}
\end{equation}
where
\begin{equation}
    \begin{aligned}
    \boldsymbol {f} &= \left[ f\left( \theta _{1},\varphi _{1}\right) ,\ldots f\left( \theta _{S},\varphi _{S}\right) \right] ^{T},\\
     \boldsymbol {c} &= \left[ C_{(0,0)} , C_{(1,-1)}, C_{(1,0)}, C_{(1,1)}, \ldots C_{(L,L)} \right] ^{T},\\
     \boldsymbol {Y} &= \left[ \boldsymbol {y}_{(0,0)} , \boldsymbol {y}_{(1,-1)}, \boldsymbol {y}_{(1,0)} \boldsymbol {y}_{(1,1)}, \ldots \boldsymbol {y}_{(L,L)} \right],\\
     \boldsymbol {y}_{lm} &= \left[ Y_{lm}\left( \theta _{1},\varphi _{1}  \right) , \ldots , Y_{lm}\left( \theta _{S},\varphi _{S}  \right)   \right] ^{T}. \end{aligned}
\label{eq: linear_eqn2}
\end{equation}
In Eq.~\eqref{eq: linear_eqn2}, $\boldsymbol{f}$ is defined for each frequency as the original magnitude values on the $S$ spatial directions, $\boldsymbol{c}$ is the desired SH coefficients, and $\boldsymbol{Y}$ contains real SH base values of up to order $L$ at the corresponding source spatial directions. To compute coefficient vector $\boldsymbol{c}$, we use the least square fit approach:
\begin{equation}
\boldsymbol{c}=\left(\boldsymbol {Y}^{T} \boldsymbol {Y}\right)^{-1} \boldsymbol {Y}^{T} \boldsymbol {f}.
\label{eq: LS_eqn}
\end{equation}

The SHT process is performed for each frequency at the truncation order $L = 7$, sufficient enough for spatial perception~\cite{Romigh2015}. The near-uniform 440-point spatial sampling scheme of the HUTUBS database supports the SHT truncation order up to 16th degree without aliasing~\cite{Zotkin2009, Rafaely2007}, and the database itself provides 16th order SH coefficients computed with complex basis. For each frequency of a subject’s HRTF, a magnitude operation is performed with considerations to the perceptual sensitivity of loudness and effective auditory filters of the cochlea~\cite{moore1995hearing}. The corresponding frequency magnitudes are sampled at 41 logarithmic frequencies in total ranging from 170 Hz to 17k Hz, and the values are converted to a dB scale. We use the real SH basis to perform the SHT on each HRTF magnitude pattern, and obtain the coefficients ($\boldsymbol{c}$ vector) of each SH base. 
By concatenating SH coefficients of each frequency bin, we obtain a lower-dimensional representation of the global HRTF pattern, which is used as the training target for the deep learning step.

Although there is no additional processing of frequency domain magnitudes, the SHT process results in a smoothed HRTF spectrum. 
Taking all frequencies and all subjects into account, the averaged SHT smoothing error accumulates to the mean and standard deviation of $(1.96 \pm 0.93)$~dB for the left ear and $(1.95 \pm 0.93)$~dB for the right ear. 
Nevertheless, SHT reconstructed HRTFs with this level of smoothing error could still produce sufficient auditory cues for spatial perception~\cite{Romigh2015, Kulkarni1998}. 
 
\begin{figure}[htbp]
    \centering
    \begin{subfigure}[b]{0.5\columnwidth}
  	\centering
  	\includegraphics[width=\textwidth]{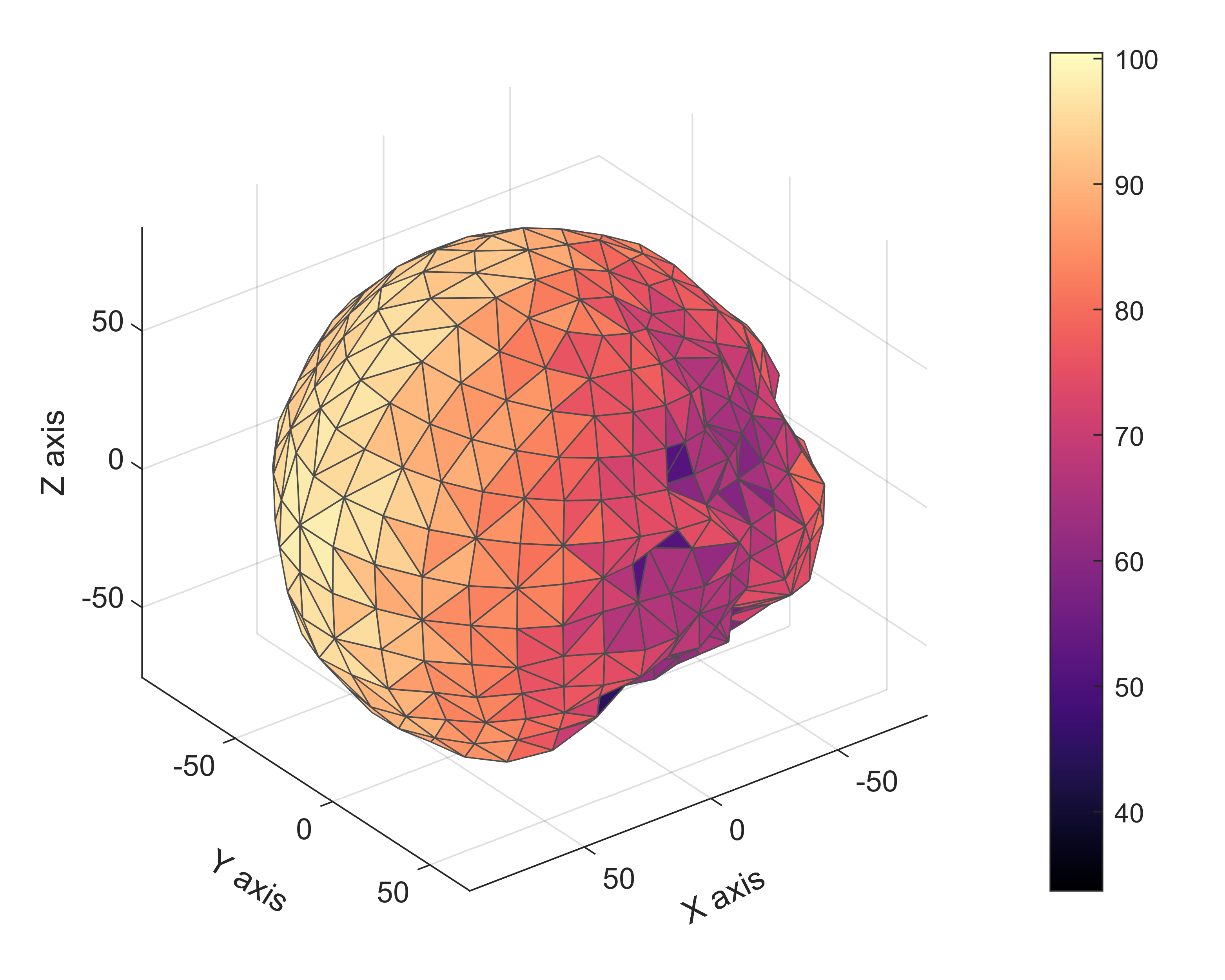}
  	\caption{}
  	\label{HRTF_SHT_4fig_first_case}
    \end{subfigure}%
    \hfill%
    \begin{subfigure}[b]{0.5\columnwidth}
        \centering
        \includegraphics[width=\textwidth]{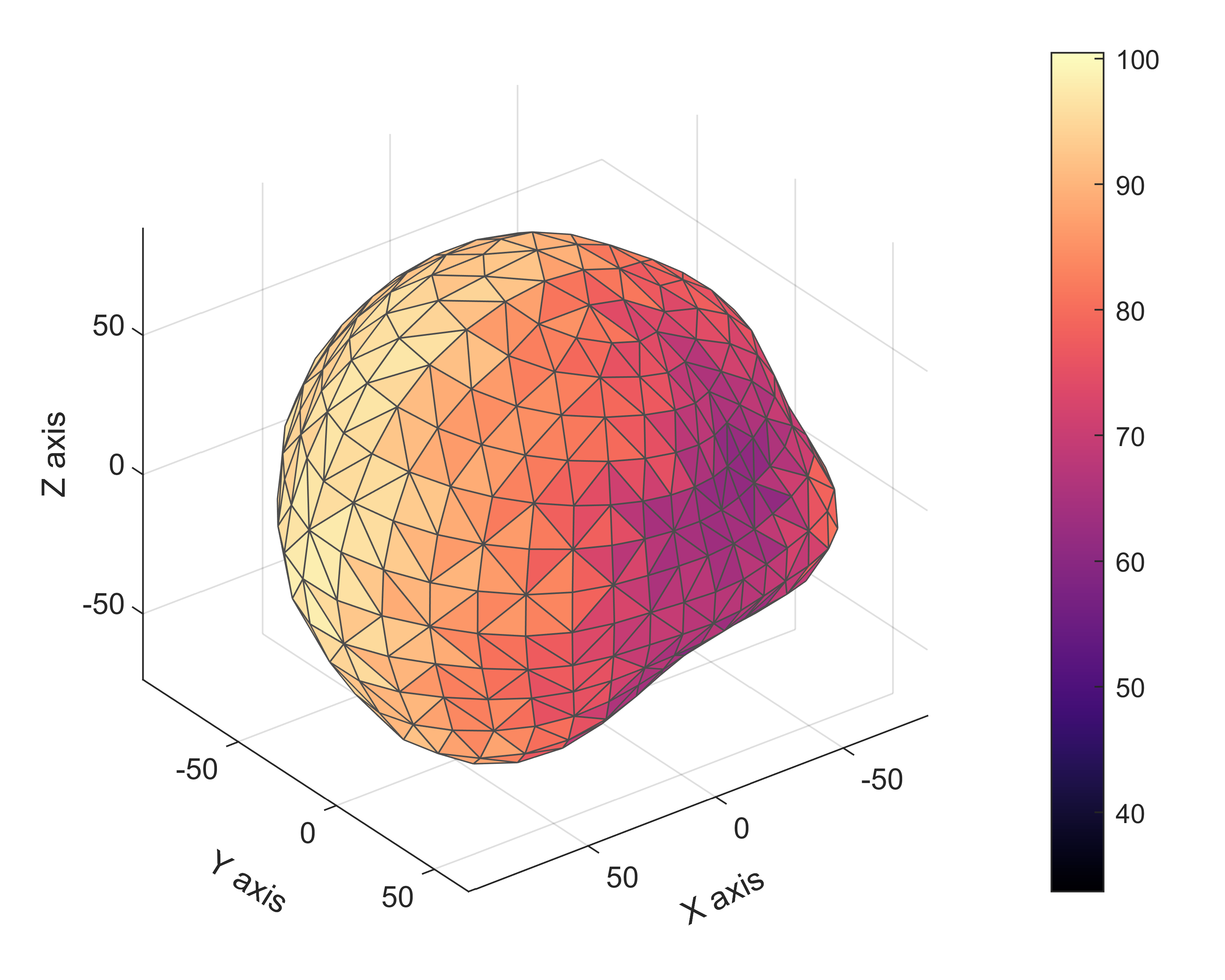}
        \caption{}
        \label{HRTF_SHT_4fig_2nd_case}
    \end{subfigure}
    \\
    \begin{subfigure}[b]{0.5\columnwidth}
        \centering
        \includegraphics[width=\textwidth]{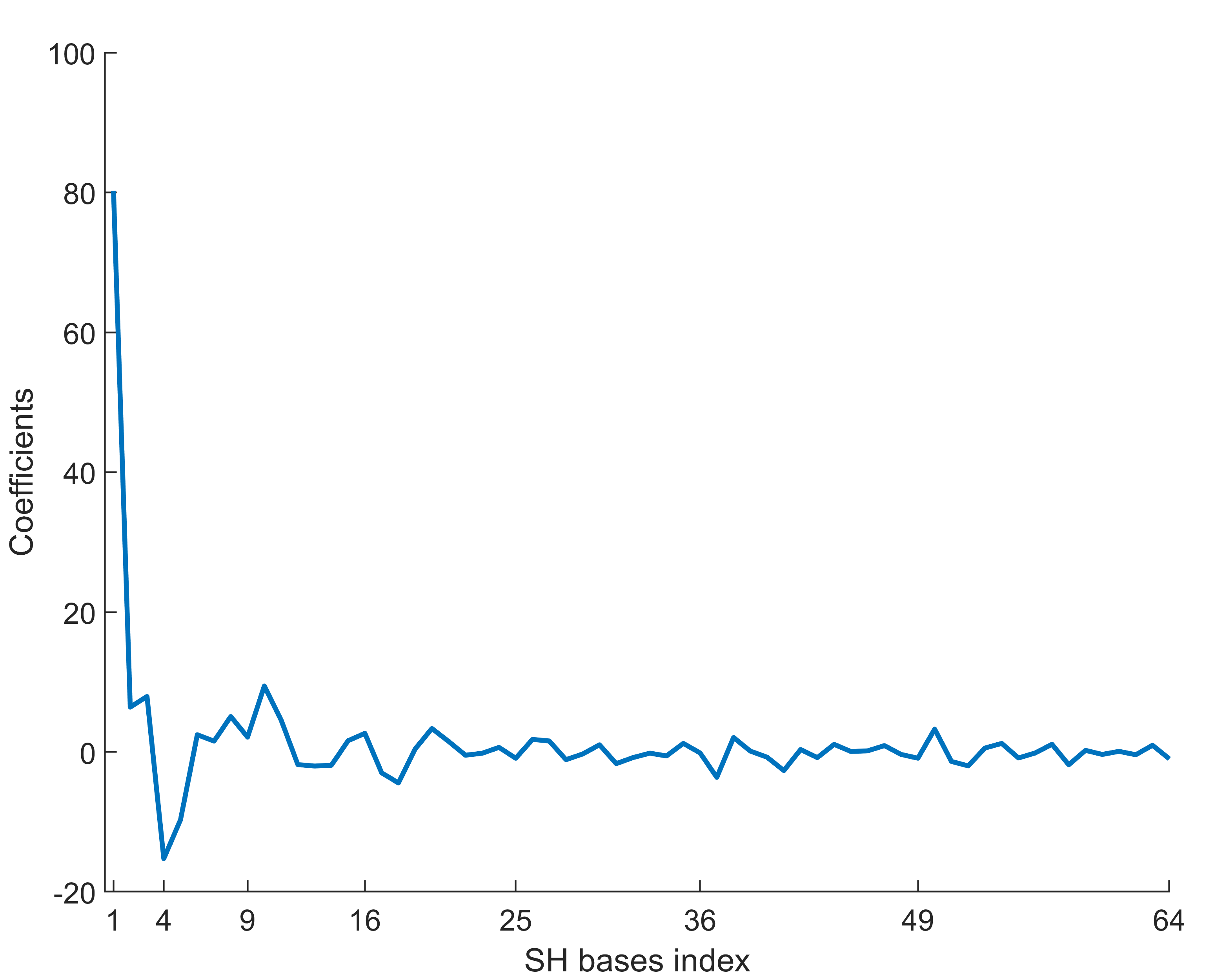}
        \caption{}
        \label{HRTF_SHT_4fig_3rd_case}
    \end{subfigure}
\subfigsCaption{Example of the HRTF magnitude pattern in dB scale and its SHT processing. (a) is the HRTF pattern at frequency of 15.9k Hz plotted in a spherical coordinate system in dB scale. The magnitude value is assigned as both the color map and the distance from each corresponding source location to the origin. (b) is the reconstructed pattern from SHT at $L =7$, and (c) shows the $(L+1)^2 = 64$ SH coefficients produced. We set the same color map in (a) and (b) for better comparison.} 
\label{fig:HRTF_SHT_4fig}
\end{figure}

An example of magnitude processing on one frequency and subject is shown in \cref{fig:HRTF_SHT_4fig}. The SHT is performed on the dB scaled HRTF magnitude pattern at a certain frequency (15.9k Hz) of one subject. The 64-d vector generated in (c) is an example of the SHT coefficients that are later used for training. 
In this example, the error produced by SHT processing is 2.89 dB averaged across all source locations, comparing the smoothed magnitudes in (b) to the original values in (a).  

\subsection{Compact representation of HRTF onsets}
Apart from the frequency domain features, the temporal features of HRTFs are also important. We choose to represent the onset time of the HRTFs, which could also be used to compute interaural time difference (ITD)~\cite{algazi2001cipic, moller1995head, andreopoulou2017identification}. The onsets are unilateral information for a set of HRTFs at either ear, indicating the time of arrival for sound at each ear across all directions. A full set of HRTFs can be reconstructed by combining the frequency part and the corresponding onsets~\cite{xie2013head}. While there are existing simplified models for HRTF temporal features such as the Woodworth formula, the finer spatial details such as front-back asymmetry are not well preserved in those methods. 

\begin{figure}[htbp]
    \centering
    \begin{subfigure}[b]{0.5\columnwidth}
  	\centering
  	\includegraphics[width=\textwidth]{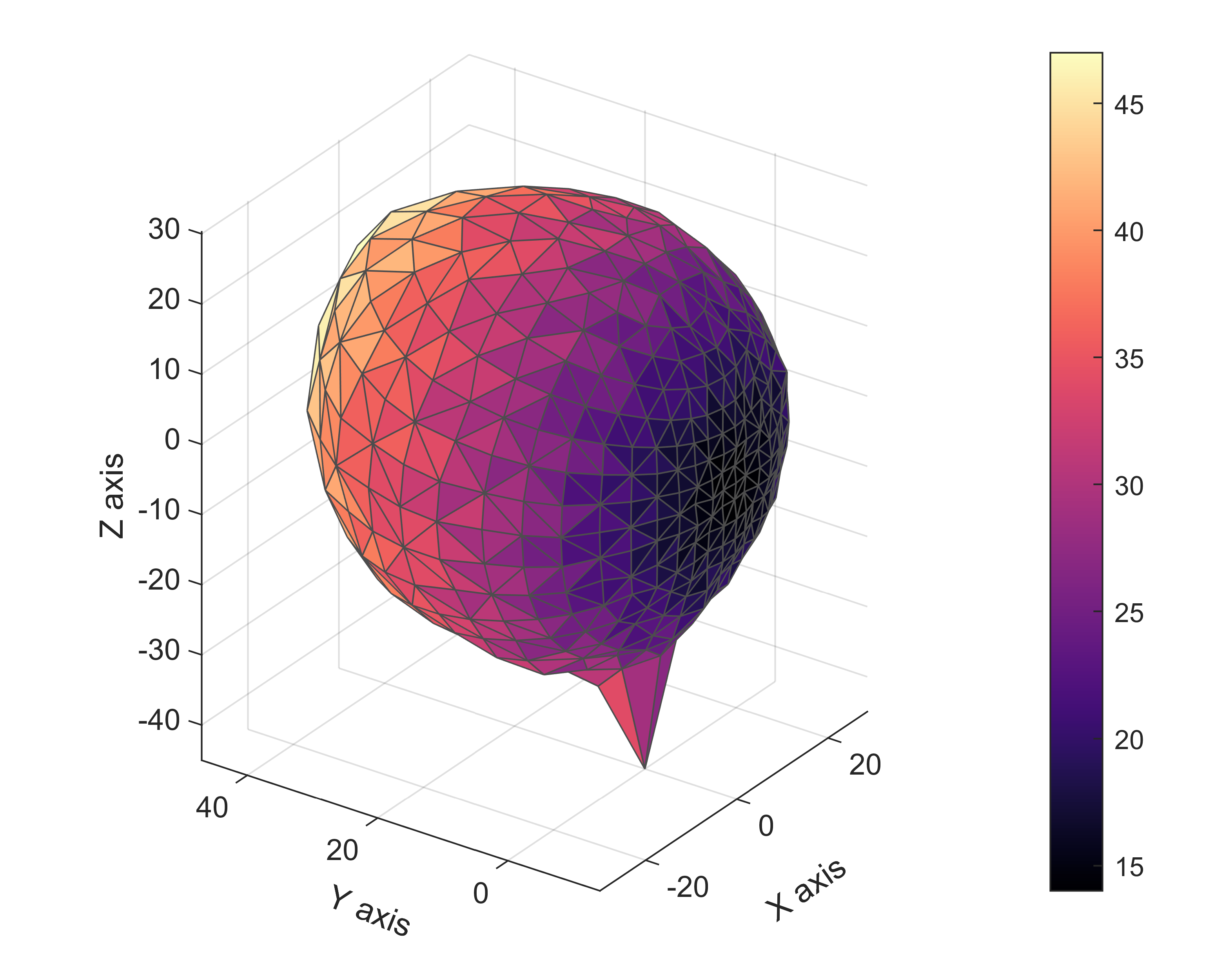}
  	\caption{}
  	\label{HRTF_onset_SHT_first_case}
    \end{subfigure}%
    \hfill%
    \begin{subfigure}[b]{0.5\columnwidth}
        \centering
        \includegraphics[width=\textwidth]{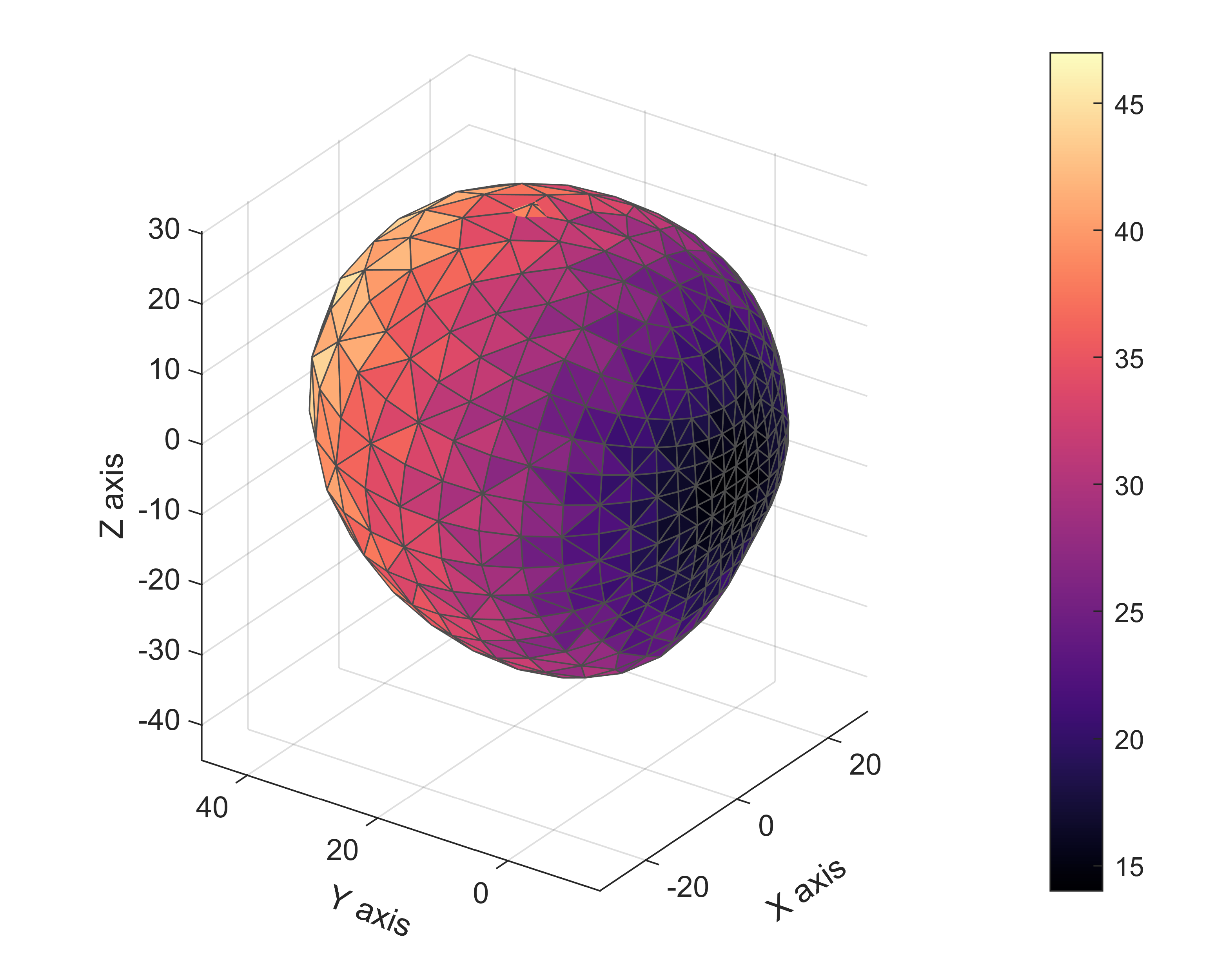}
        \caption{}
        \label{HRTF_onset_SHT_2nd_case}
    \end{subfigure}
    \\
    \begin{subfigure}[b]{0.5\columnwidth}
        \centering
        \includegraphics[width=\textwidth]{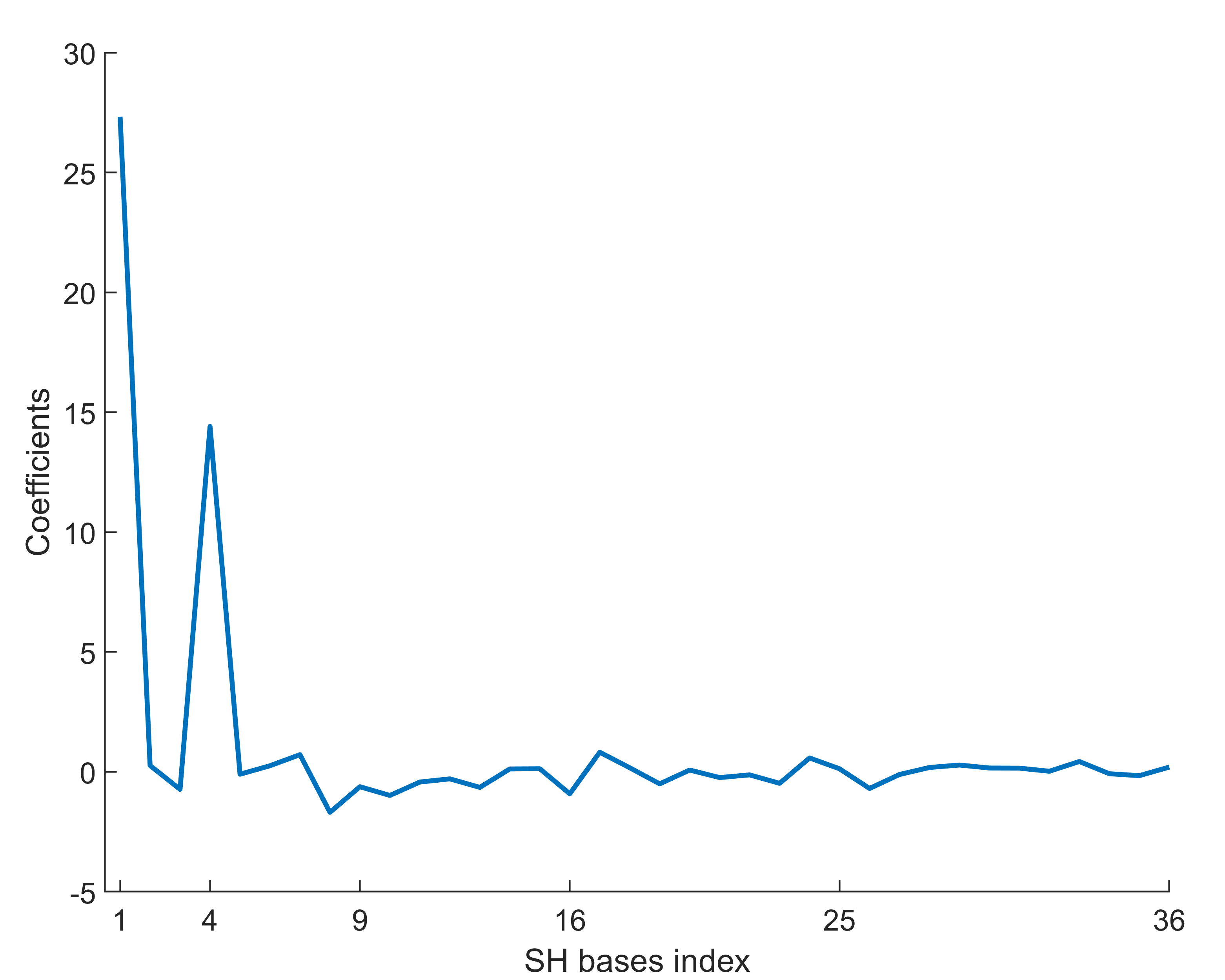}
        \caption{}
        \label{HRTF_onset_SHT_3rd_case}
    \end{subfigure}

\subfigsCaption{Example of the HRTF onset pattern and its SHT processing. (a) is the onset pattern plotted in a spherical coordinate system. The onset value is assigned as both the color map and the distance from each corresponding source location to the origin. (b) is the reconstructed pattern from SHT at $L =5$, and (c) shows the $(L+1)^2 = 36$ SH coefficients produced. We set the same color map in (a) and (b) for better comparison.}
\label{fig:HRTF_onset_SHT_4fig}
\end{figure}

In this work, SHT-based processing methods are also applied to capture the global feature of the onset times in HRTFs. We first compute the onset time for each location, and then convert them with the same SHT scheme into SH coefficients using Eq.~\eqref{eq: LS_eqn}, where $\boldsymbol{f}$ contains these onset times in this process. 
Due to its relatively simple pattern, the truncation order is set to 5, producing 36 coefficients for each ear's onset. These coefficients are later used as the ground truth for another neural network that performs onset prediction.

In \cref{fig:HRTF_onset_SHT_4fig}, the global onset SHT processing is demonstrated. The onset values are smaller on the ipsilateral side and higher on the contralateral side, due to the off-center position of the ears. The order of the positions in HUTUBS database are starting from the top (+90 degree elevation) and going through a complete circle at each elevation until down to –90 degree elevation. 
Note there is one particular large onset value located at the lowest location in (a). 
This is resulted from the source location design of the HUTUBS database, which includes one location directly beneath the subject. This outlier has little influence on the SHT processing, but we still exclude it in the evaluation step for the onset prediction performance. Each subject's HRTF onset data produces a 36$\times$2 coefficient matrix considering the left and right ears. These coefficients are used as the learning targets for the onset prediction.

\section{Compact representation of mesh geometry}
\label{sec: compact_represent_mesh}

To efficiently represent features embedded in the head geometry and link them to acoustic features in HRTFs, some processing techniques are necessary. Here we adapt the common processing approach used in computer graphics to achieve a more compact representation of the subjects' head geometry.

\begin{figure}[t!]
    \centering
    \begin{subfigure}[b]{0.7\columnwidth}
        \includegraphics[width=\textwidth]{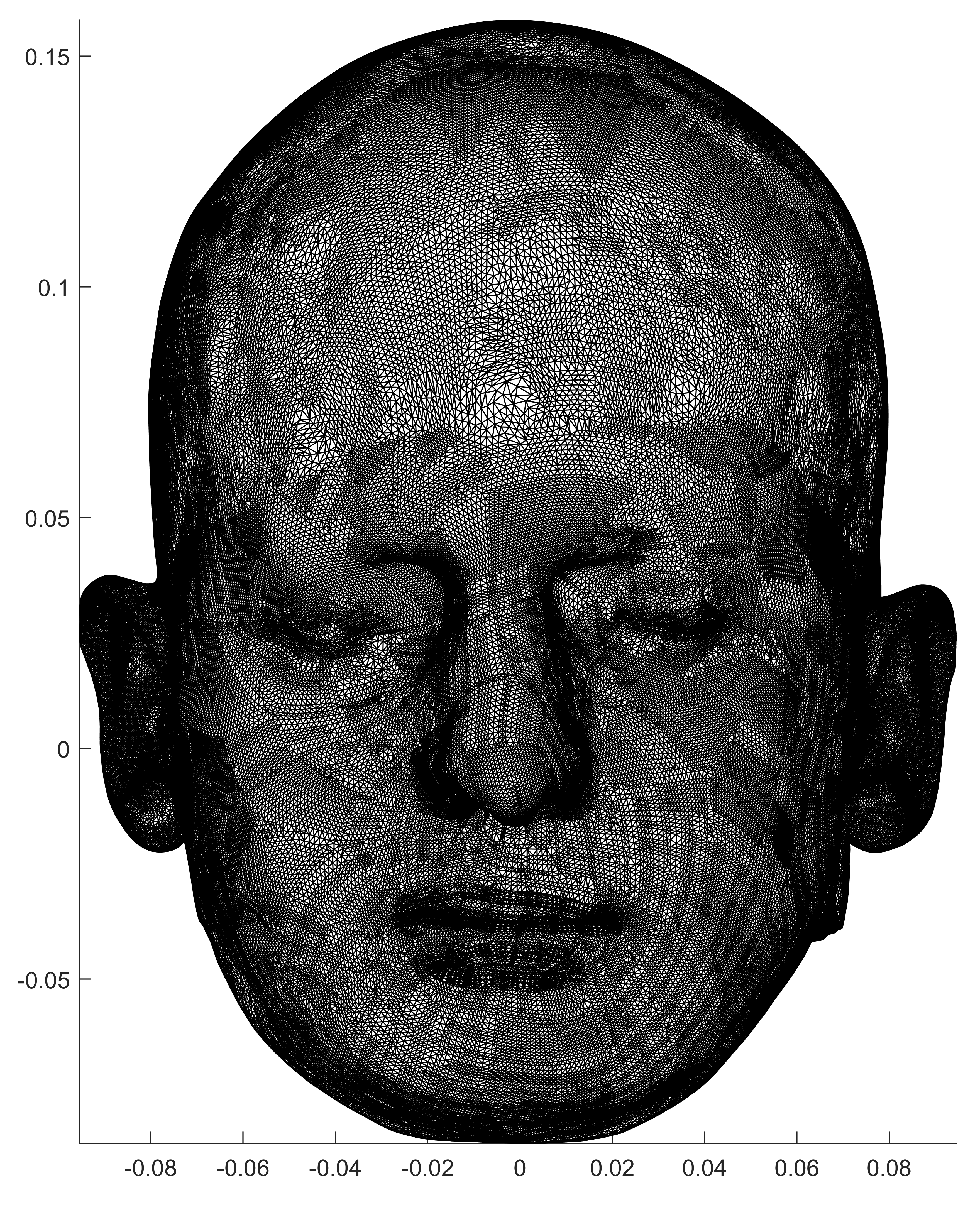}
        \caption{}
    \end{subfigure}
    \begin{subfigure}[b]{0.7\columnwidth}
        \includegraphics[width=\textwidth]{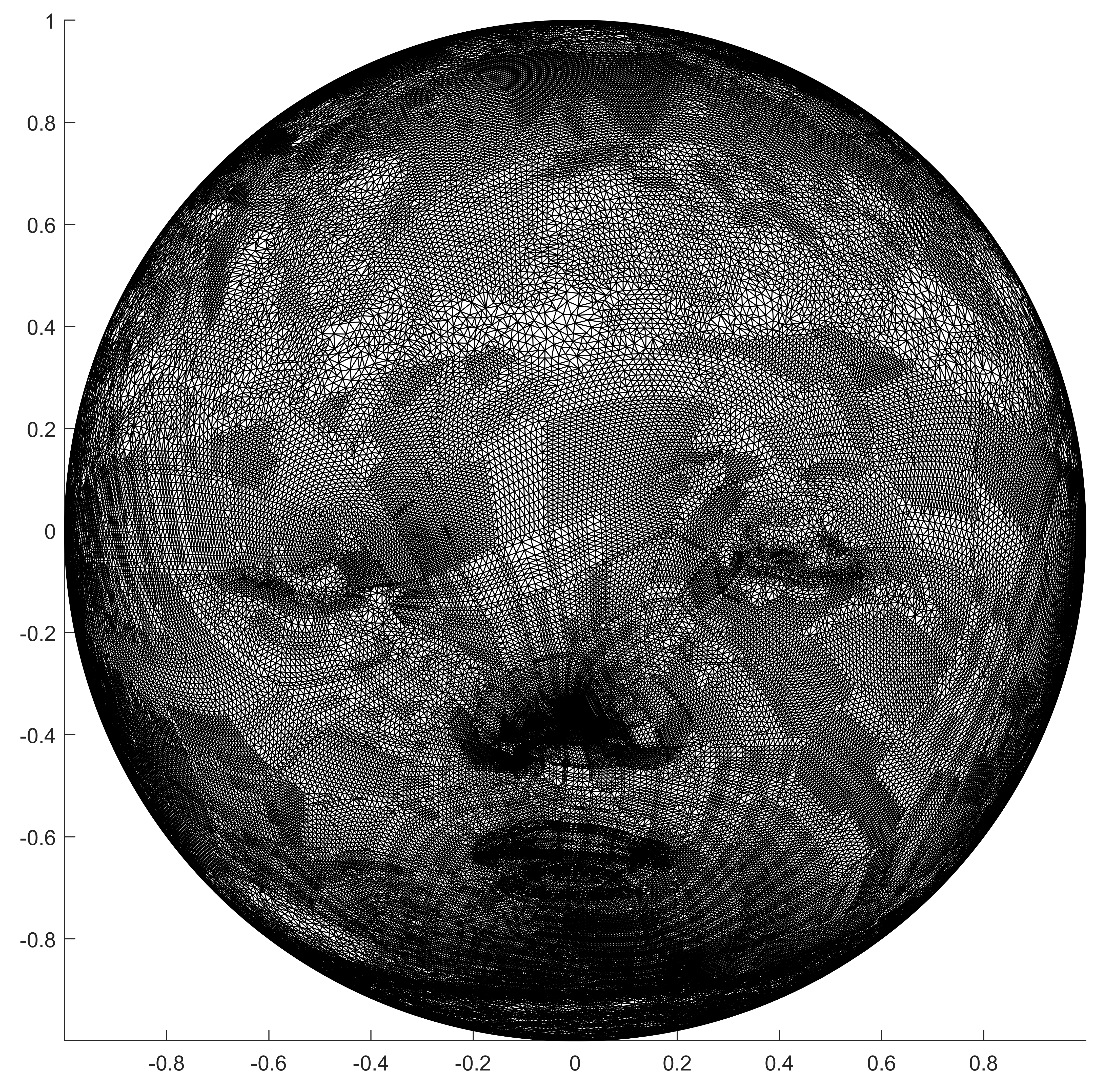}
        \caption{}
    \end{subfigure}


\subfigsCaption{Example of head mesh mapping process. (a) is a subject's original head scan, (b) is the corresponding conformal mapping onto the unit sphere.}
\label{fig:head_and_conformalMap}
\end{figure}

\subsection{Mesh parameterization}
\label{ssec:mesh_para}

To model a non-convex surface geometry such as a human's head, a certain surface mapping process is desired to project the original mesh into a convex space.
For the surface of a closed 3D object, a bijective mapping can be applied such that its geometry can be represented by a radial function $r(\theta, \varphi)$ in the new mapping space \cite{Shen2009}. This mapping process is called mesh parameterization, which refers to the process of mapping a 3D triangulated mesh onto a planar domain. Most mesh parameterization algorithms are based on the theories in differential geometry. Here we utilize the existing parameterization method introduced in a previous work \cite{choi2015flash}, and perform conformal mapping to project the subjects head mesh onto a unit sphere surface, shown in Figure \ref{fig:head_and_conformalMap}. In the spherical mapping space, each vertex location is jointly represented by three functions, $x(\theta, \varphi)$, $y(\theta, \varphi)$ and $z(\theta, \varphi)$, which are orthogonal to each other. Through this step, we are able to obtain a regularized sampling space to perform further analysis. The mesh parameterization process is still feasible when a complete head scan is unavailable, in case some datasets only include scans for the ear region. The stand-alone ear scans can be attached to an ellipsoid with the same width/height/depth dimensions as the subject’s head.

With each head mesh mapped onto a unit sphere, we thereby normalize each head as spheres with a radius of 1. The resulting coefficients of each SCH basis are therefore corresponding to the geometry on a unit sphere, rather than the actual scale of each head. Correspondingly, HRTFs should be stretched/compressed in the frequency domain according to each subject's head dimensions. We calculate each subject's equivalent head radius and generate the normalization factor for everyone based on Subject 1 in the database. The frequency compensation is then applied to each subject's HRTFs according to the normalization factor.
After the compensation, the machine learning framework is able to map from the normalized geometric features to the normalized acoustic features.

\subsection{Spherical cap harmonics analysis for ear geometry}
\label{ssec:scha}
Due to the variety and complexity of ear shapes, it is meaningful to focus on the ear regions that contribute most to the spectral cues~\cite{xie2013head}. To efficiently extract geometry features within a certain region, we employ spherical cap harmonics analysis (SCHA)  to model the ears.

\begin{figure}[]
\centering
\includegraphics[width=\columnwidth]{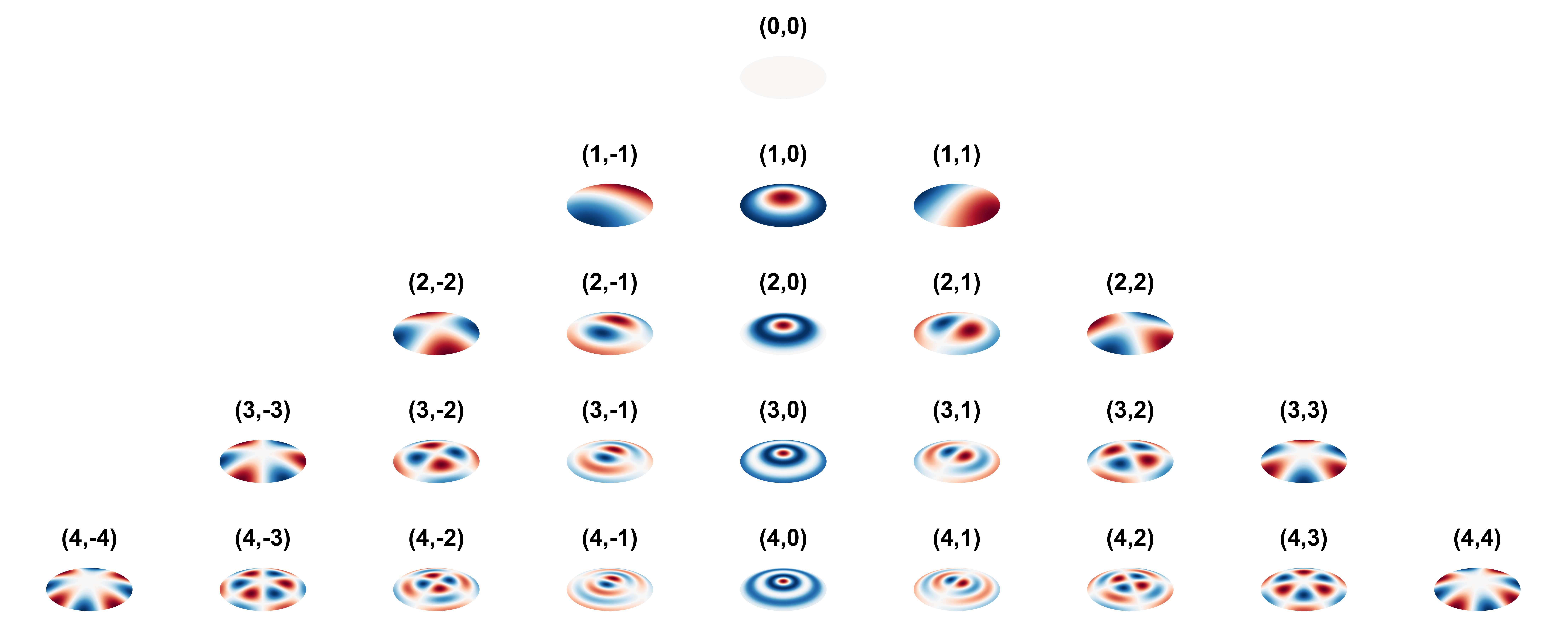}
\caption{SCH bases up to $k = 4$ at half cone angle of 25 degrees. Numbers in parenthesis are degree index $k$ and order $m$, where $-k \leqslant m \leqslant k$.}
\label{fig:SCH_bases}
\end{figure}

The SCH is a set of orthogonal basis defined on a spherical cap of a certain cone angle. The complex-valued SCHs of degree $l$ and order $m$ for a cap of size $\theta_{c}$ can be expressed as~\cite{Shaqfa2021}:
\begin{equation}
{ }^{\theta_{c}} C_{k}^{m}(\theta, \phi)=\bar{P}_{l(m)_{k}}^{m}(\cos \theta) e^{i m \phi},
\label{eq:SCH_base}
\end{equation}
where $\bar{P}_{l(m)_{k}}^{m}(\cos \theta)$ is the associated Legendre function (ALF) defined within the latitude region (polar cap), and $k$ is the degree index.
Similar to the SH basis layout, the SCH bases up to the 5th degree at the half-cone angle of 25 degrees are shown in \cref{fig:SCH_bases}. Note that in practice, we compute SHCA using the real-valued SCH basis, similar to the previous SHT step, which simplifies the machine learning design while maintaining orthogonality.

A 2D signal $f(\theta, \varphi)$ of angles $\theta$ and $\varphi$ can be expanded using SCH basis as~\cite{Haines1985}: 
\begin{equation}
f(\theta, \phi)=\sum_{k=0}^{\infty} \sum_{m=-k}^{k} { }^{\theta_{c}} q_{k}^{m} \cdot { }^{\theta_{c}} C_{k}^{m}(\theta, \phi),
\label{eq:SCH_expansion}
\end{equation}
where ${ }^{\theta_{c}} q_{k}^{m}$ are the SCH coefficients. In our practice, we use the same least-square fitting approach to compute these coefficients of each SCH basis, only changing the basis and sample values in Eq.~\eqref{eq: LS_eqn} according to the sphere cap case.

\begin{figure}[t!]
    \centering
    \begin{subfigure}[b]{0.49\columnwidth}
  	\centering
  	\includegraphics[width=\textwidth]{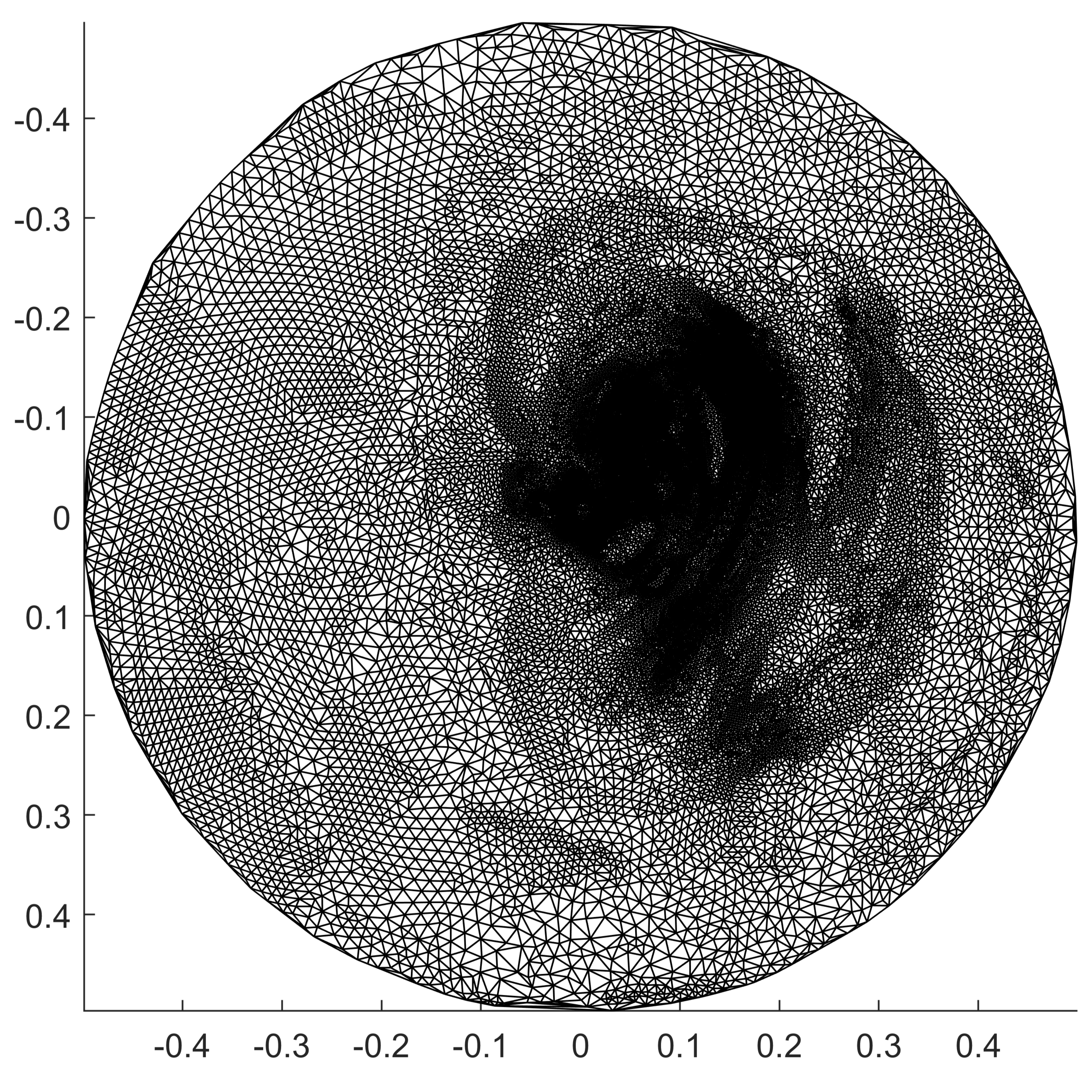}
  	\caption{}
  	\label{ear_sch_demo_first_case}
    \end{subfigure}%
    \hfill%
    \begin{subfigure}[b]{0.49\columnwidth}
        \centering
        \includegraphics[width=\textwidth]{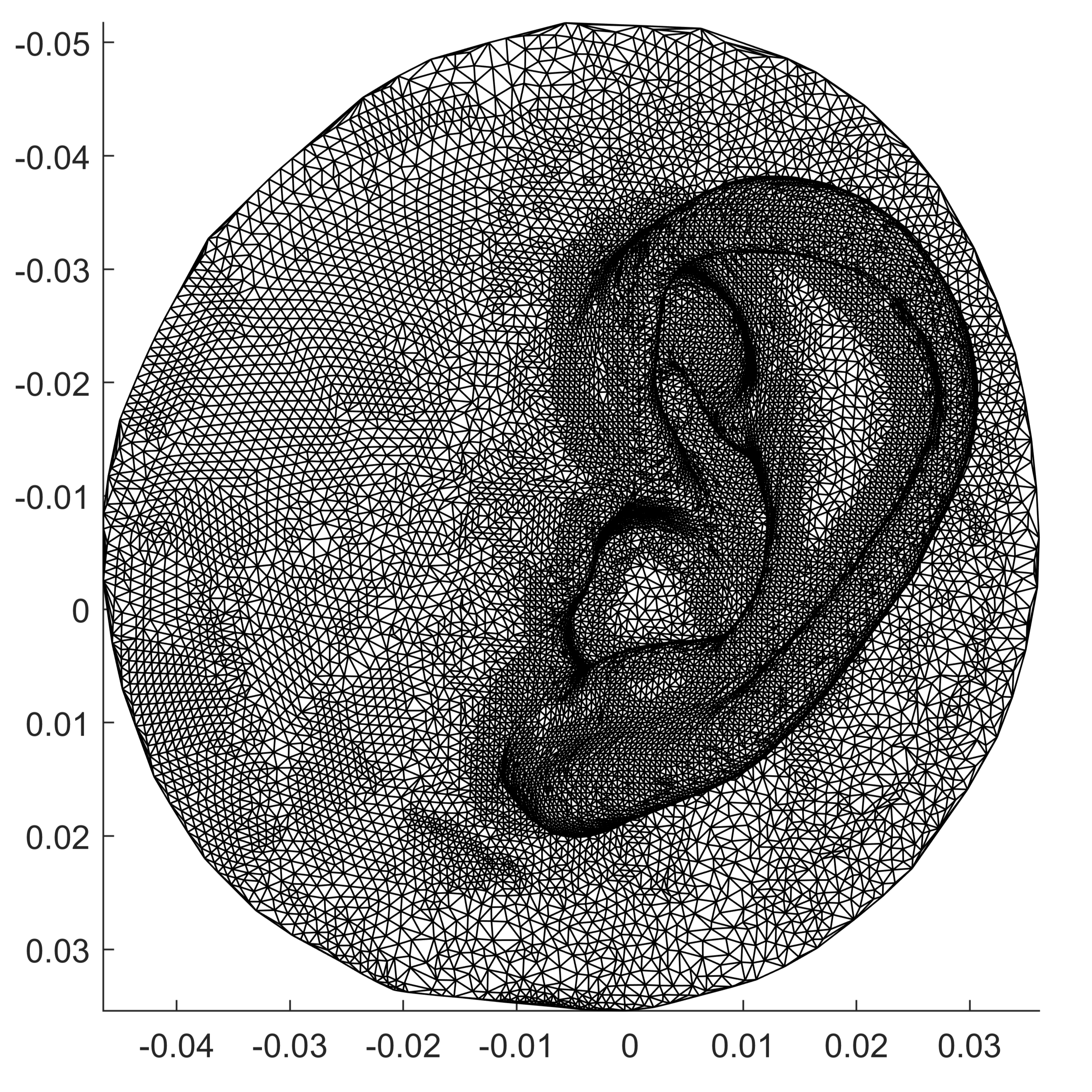}
        \caption{}
        \label{ear_sch_demo_2nd_case}
    \end{subfigure}

    \begin{subfigure}[b]{0.49\columnwidth}
        \centering
        \includegraphics[width=\textwidth]{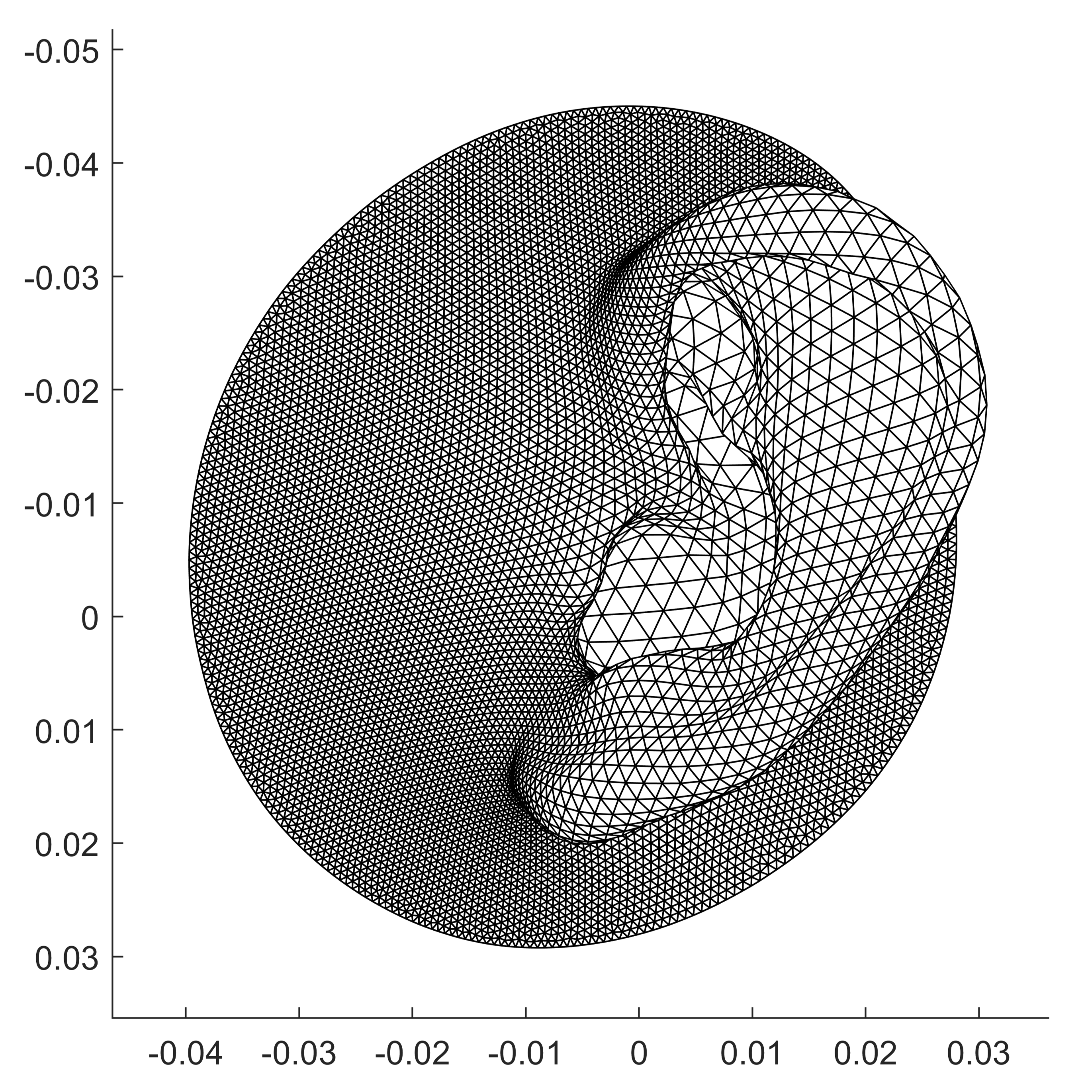}
        \caption{}
        \label{ear_sch_demo_3rd_case}
    \end{subfigure}
    \hfill
    \begin{subfigure}[b]{0.49\columnwidth}
        \centering
        \includegraphics[width=\textwidth]{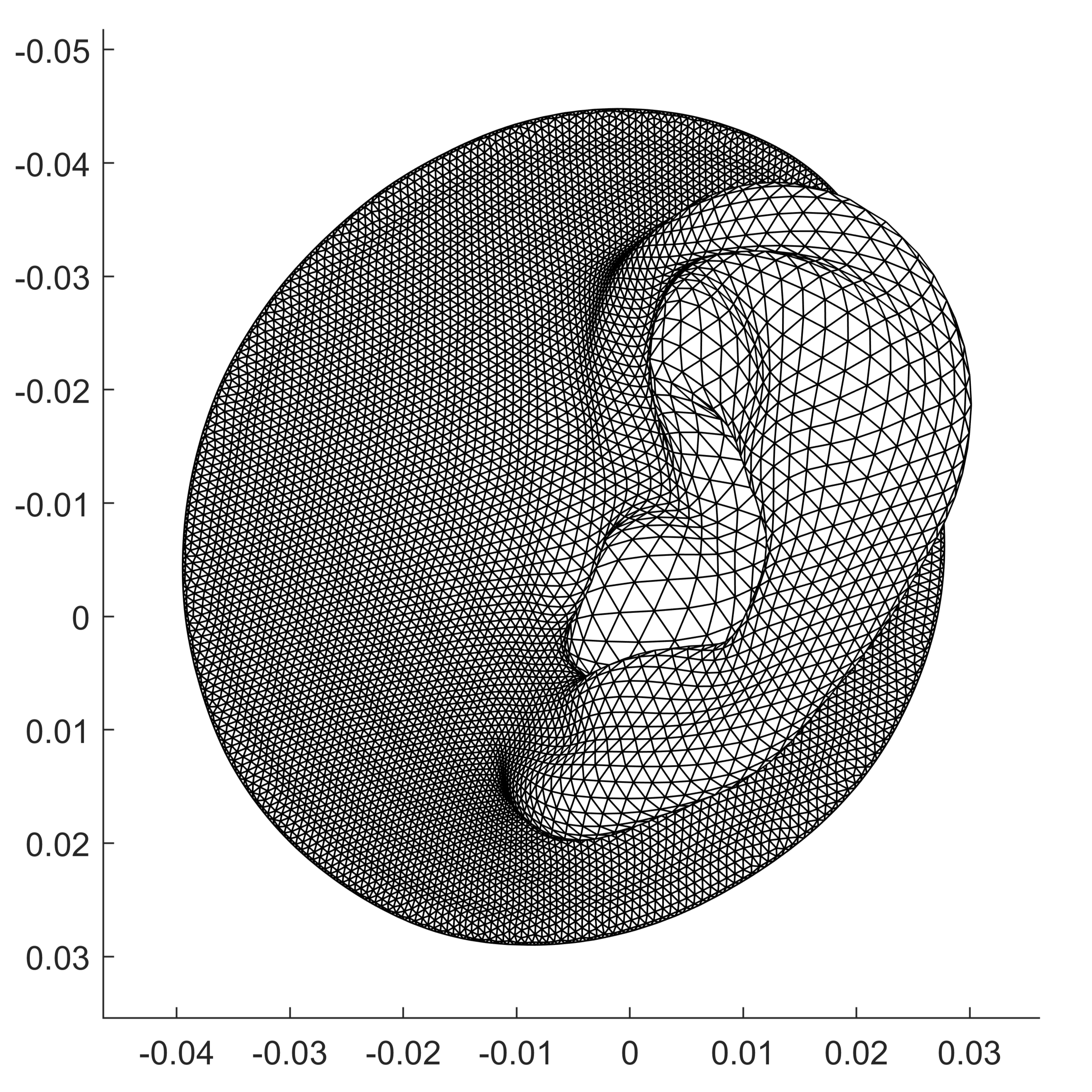}
        \caption{}
        \label{ear_sch_demo_4th_case}
    \end{subfigure}    
    
\subfigsCaption{Ear SCHA process flow. (a) is the cap area cropped from head spherical mapping, with the ear regions within. The cap is centered at the ear canal entrance, with a half-cone angle of 30 degrees. (b) is the corresponding ear mesh in the original mesh space. (c) is the remeshed ear with uniform cap sampling (9062 vertices) with a half cone angle of 25 degrees. The SCHA is then applied with these 9062 vertex samples. (d) is the reconstructed mesh after the SCHA at a truncation order of 20. 
Meshes in (b) and (c) are set to the same scale for better comparison.} 
\label{fig:ear_sch_demo}
\end{figure}

We develop a processing flow for each subject's ear mesh: (1) First, we isolate the ear region in a cap area. The caps are cropped from the head mesh spherical mapping from the previous step, and the center of each cap is set at the entrance of each ear canal, the location where HRTFs are measured. The half-cone angle is set to 30 degrees, ensuring the inclusion of the ear region for all subjects while excluding the unnecessary area as much as possible. (2) The ear geometry is then remeshed according to a uniform sampling scheme (9062 vertices) on a slightly smaller cap of 25 degrees half-cone angle \cite{Roca2010}, to facilitate the SCH calculation. Similar to the SHT case, the SCHA also desires uniform sampling within the sampling space. The remesh cone (25 degrees) is slightly smaller than the cropped cone (30 degrees), to ensure the neighboring reference vertices for the remesh process. (3) After the remeshing, the SCHA process is applied to each $x$, $y$, and $z$ dimension. The coefficients are computed similarly to the SHT, with the least-square fit of the SCH basis matrix and the original value vector. 
The whole process is demonstrated in \cref{fig:ear_sch_demo}.

\begin{figure*}[htbp]
    \centering

    
    \subfloat[]{\includegraphics[width=0.65\columnwidth]{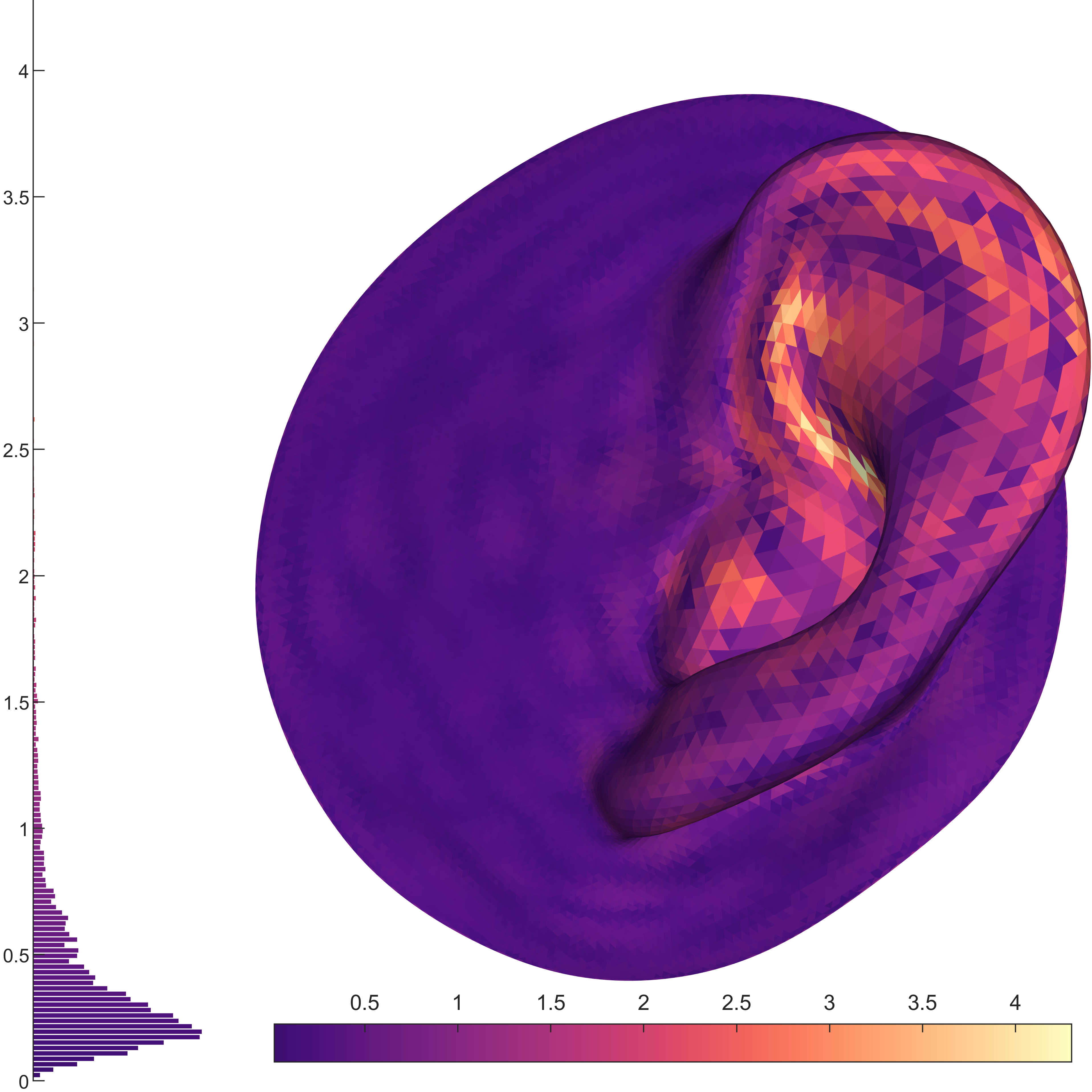}}%
    \subfloat[]{\includegraphics[width=0.65\columnwidth]{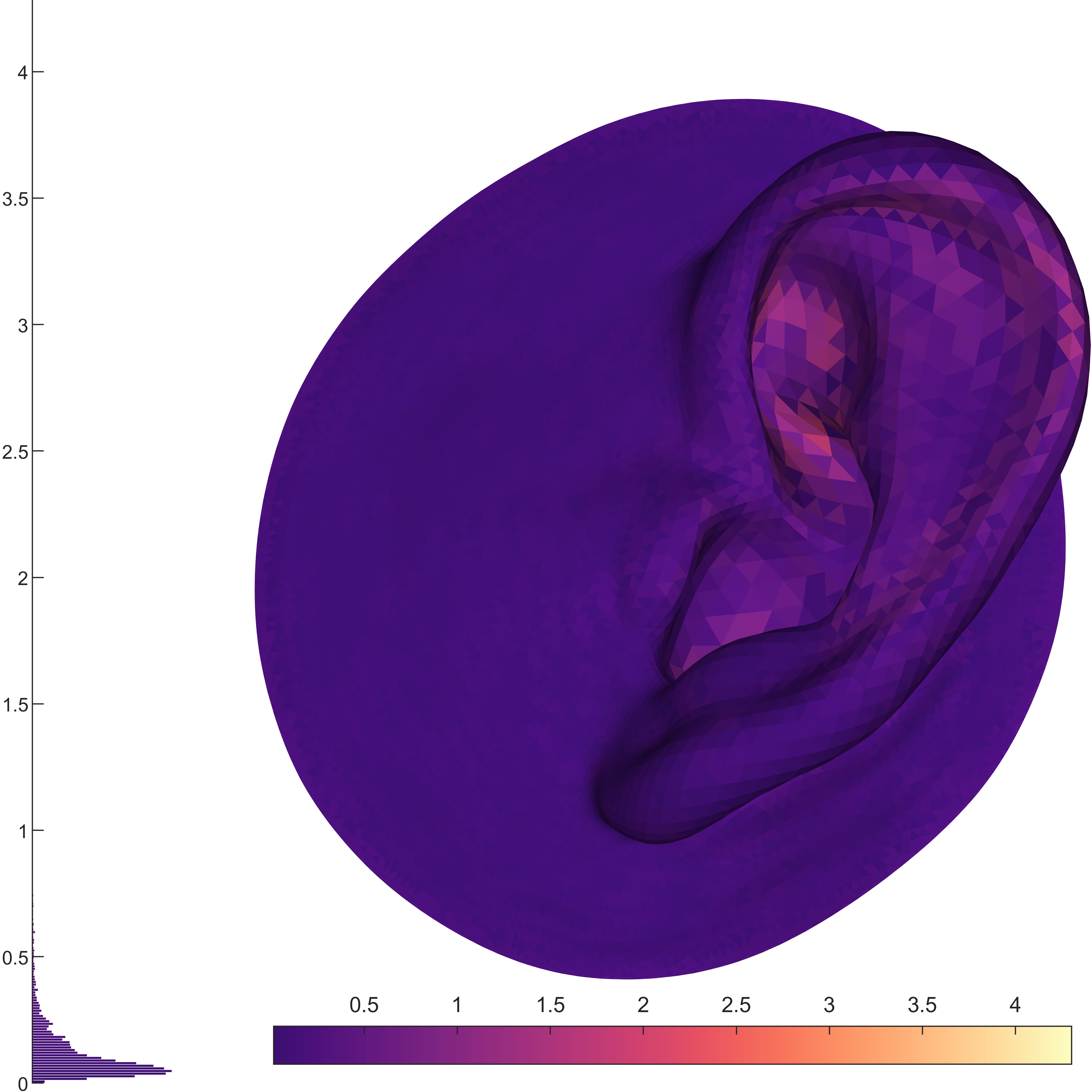}}%
    \subfloat[]{\includegraphics[width=0.65\columnwidth]{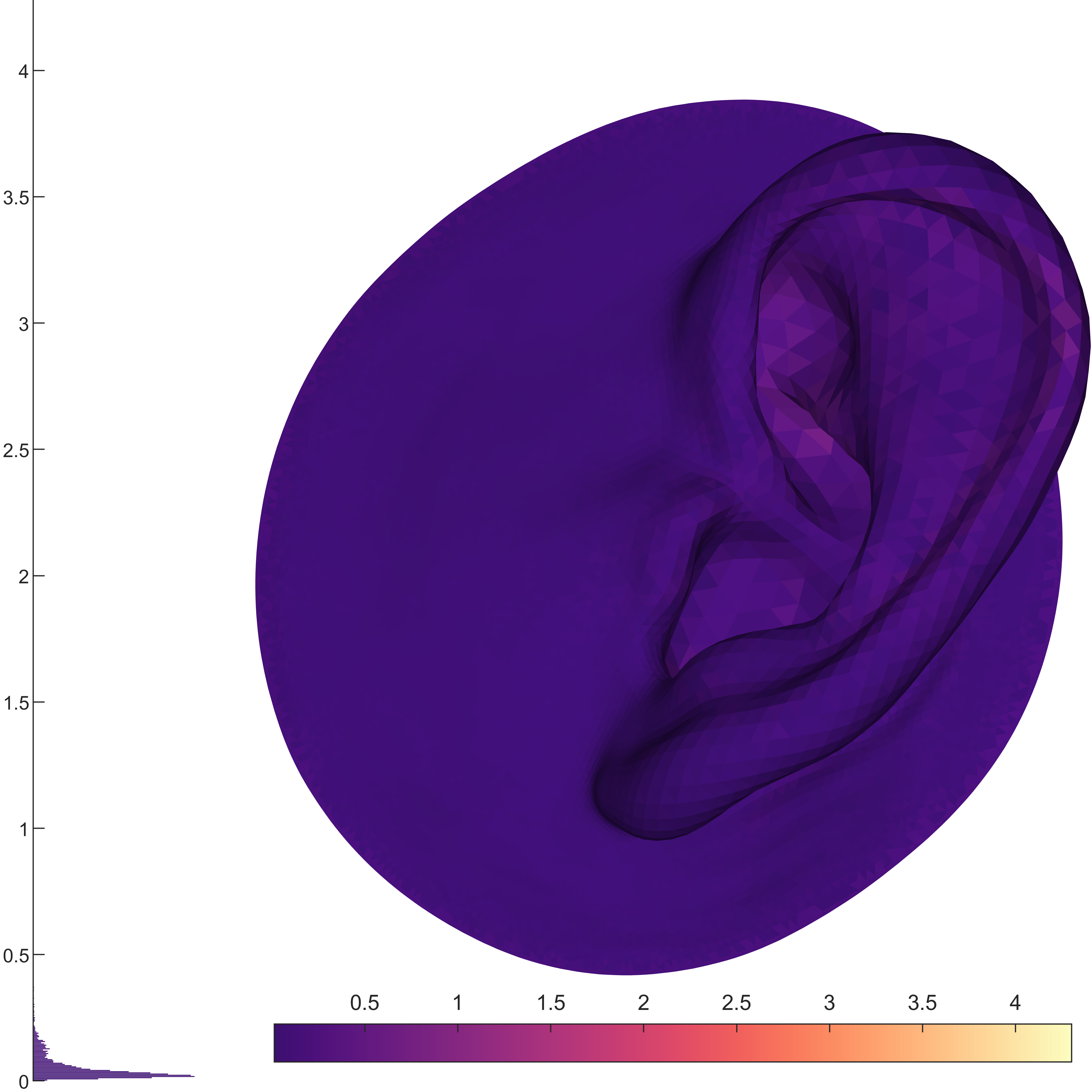}}%

\caption{Comparison of the SCH reconstruction error distance (in mm) at different truncation orders visualized on the ear mesh. (a) 10th degree. (b) 20th degree. (c) 30th degree. For each panel, the error distribution is plotted on the left side. We set the range of all the distributions and the color map according to the 10th degree case for better comparison.}
\label{fig:ear_sch_recons_comparison}
\end{figure*}

The HUTUBS database has 96 subjects in total, 53 of which have valid mesh scans. With 53 subjects with valid head mesh scans, we perform SCHA to extract the ear features for all of them.
We evaluated the SCH reconstruction error on the ear meshes to determine the optimal truncation order of SCHA.
In \cref{fig:ear_sch_recons_comparison}, we show the comparison of the 10th, 20th, and 30th degree SCH reconstruction error distances (in mm) of an ear mesh (HUTUBS subject 1). The error distance values are calculated as the Euclidean distance between the position of the vertex in the remeshed ear (shown in Figure~\ref{fig:ear_sch_demo}c) and the corresponding vertex in the SCH reconstructed ear.
For the 10th degree case, the maximum error is higher than 4 mm, while most errors are within 1 mm. For 20th degree case, the maximum error occurs around 1.5 mm, while most errors are below 0.5 mm. For the 30th degree case, the errors are sub-millimeter level for all vertices, with most of them occurring within 0.2 mm.

The medians of the 20th SCH reconstruction errors across all vertices are $(0.081,0.076)$ mm for left and right ears for HUTUBS subject 1, with mean values and standard deviations of $(0.123 \pm 0.135)$ mm and $(0.116 \pm 0.124)$ mm for the left and right ears.
Taking all the subjects into account, the error mean and standard deviation are $(0.143 \pm 0.116)$ mm for left ears, and $(0.146 \pm 0.117)$ mm for right ears. These error values are way below the threshold (less than 1mm) for meshes yielding perceptually valid HRTFs~\cite{Ziegelwanger2015, Pollack2021}, indicating that the SCH reconstruction errors are negligible between the reconstructed and the original ear meshes.
 
\begin{figure}[htbp]
    \centering
    \includegraphics[width=0.9\columnwidth]{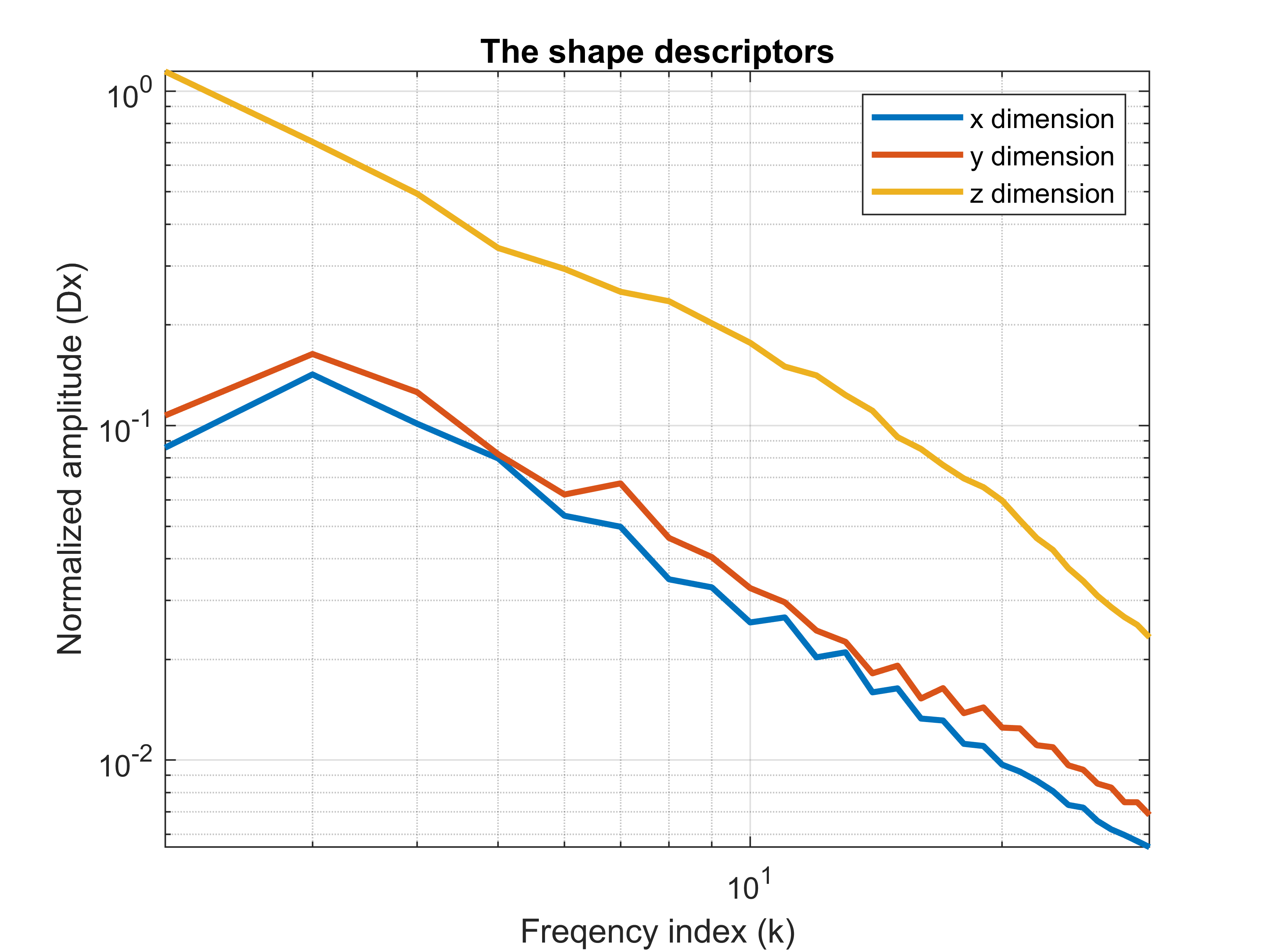}%
\caption{Ear SCH power analysis. The SCH amplitude (also known as shape descriptors) are plotted as a function of spatial frequency index $k$ for $x$, $y$, and $z$ dimension, respectively. The results are averaged across all subjects' meshes. }
\label{fig:ear_sch_energy}
\end{figure}

Furthermore, we perform power spectral analysis for the SCH coefficients, and the averaged SCH power of 3 dimensions(x,y,z) are shown in \cref{fig:ear_sch_energy} where SCHA is performed with a truncation order up to 30, with all subject's results included. In this figure, we plot the combined amplitude of SCH coefficients of the same degree, as a function of degree index $k$. This indicates the accumulated power at a certain wavelength (referred to as `shape descriptors' in related work \cite{kazhdan2003rotation, Zhao2017}) within the cap area. As shown in the figure, the SCH power decays significantly after the 20th degree. Across all subjects, 20th-degree SCH representations cover more than 95\% energy compared to the case of the 30th degree, for both left and right ears.

Therefore, the truncation order is set to 20. Compared to the original mesh, these SCH coefficients manage to use only less than 0.5\% amount of the original data while maintaining most of the information of the ear geometry. A relatively low truncation order value would also benefit the machine learning design, which reduces overfitting and the complexity of the network.

With degree index $k$ set as 20, the SCHA produces a 441$\times$3 coefficient matrix for each ear, which are sent as the input for the neural network training. The head and torso's larger-scale dimensions such as head height/width and shoulder width are sent into the network as well, since they are not included in the ear meshes.

\section{Deep learning design and implementation}
\label{sec:Deep_learning_design}

\begin{figure}[htbp]
    \centering
    \includegraphics[width=\columnwidth]{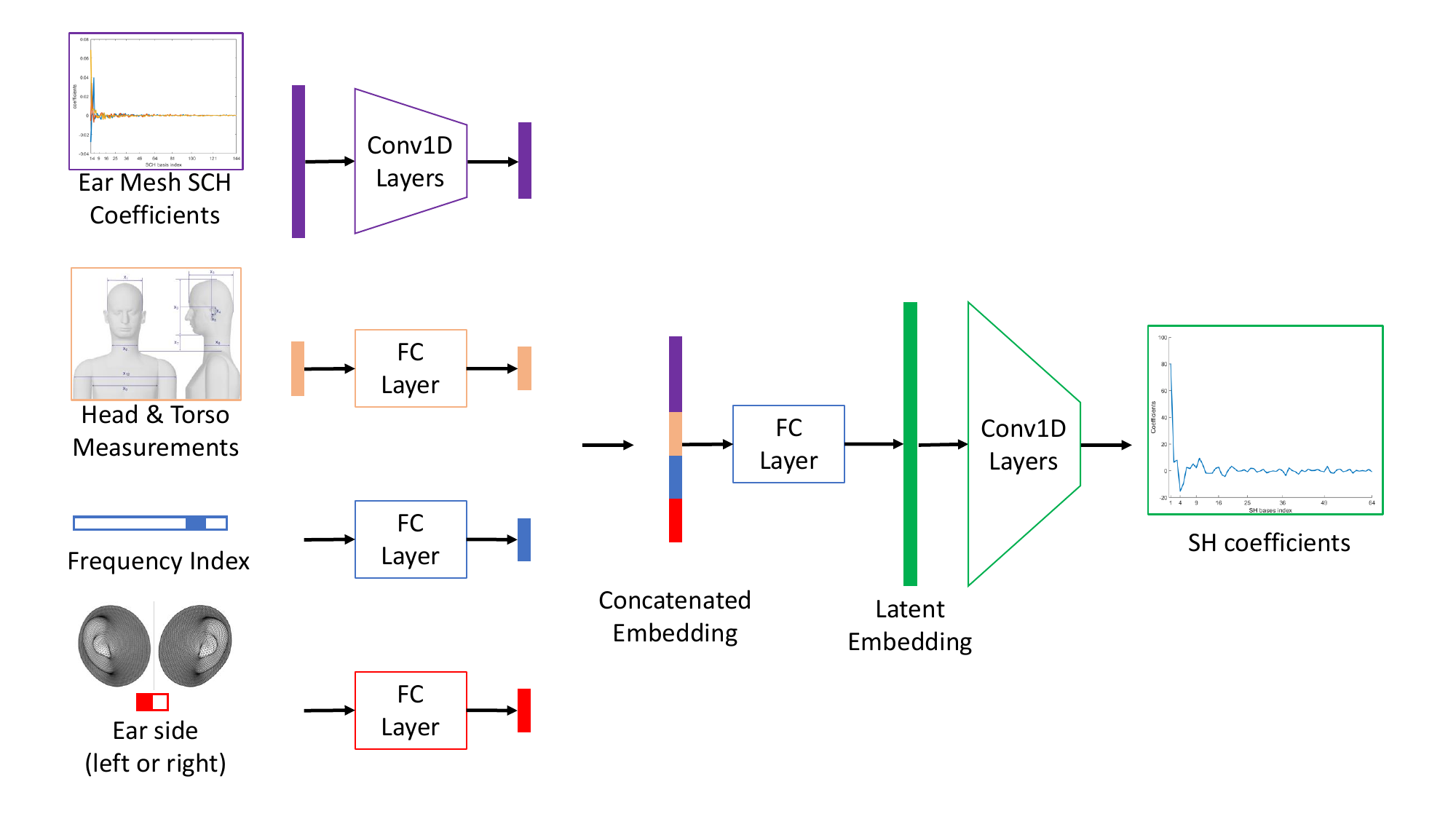}
    \caption{Diagram of the deep learning neural network design. Our input is subjects' geometry information, including ear mesh SCH coefficients for the xyz dimension, head and torso measurements, one hot encoding for the frequency index and ear side. The system would output the SH coefficient as HRTFs' compact representations.}
    \label{fig:NN_diagram}
\end{figure}

A deep learning-based model is designed to map the information of the scanned head mesh to the SH representation of the HRTF. The model structure is illustrated in \cref{fig:NN_diagram}. The input to the neural network is the SCH of the ear mesh, head and torso measurements, frequency, and ear side (left/right).
As the design choice of the neural network, the learning input and the target are both compact representations of the mesh and the HRTFs; This allows us to efficiently link the connections between the intrinsic features of the geometry and the acoustics measurements.

The ear SCH coefficients calculated from the previous step in Section~\ref{ssec:scha} are processed through a convolutional neural network (CNN) to encode this information. The CNN consists of 6 layers of 1D CNN that transforms the input dimension of 441 to the output embedding vector with a dimension of 64. The other macro-scale measurements such as head height/width and torso width (corresponding to $x_1$ through $x_{17}$ in the HUTUBS anthropometric measures\cite{brinkmann2019hutubs}) are also fed into fully connected (FC) layers to compute larger-scale features of the geometry. For a total of 13 head/torso measurements, a FC layer transforms the 13-d vector into 32 dim. We consider the frequency and ear (left or right) as side information and feed them into an FC layer to obtain their embedding respectively. The one-hot frequency input is transformed into a 16-d embedding vector, and the left or right information is encoded into another 16-d embedding vector to keep balance with other information. We then concatenate these embedding outputs and use another FC layer to fuse the information. The concatenated embedding is hence with dimension 128, and the FC layer transforms it into a 256-d vector. With the following 5 layers of 1D CNN, the 64-d SH coefficients of HRTF are predicted. 
The training loss is calculated with the mean square error (MSE) of the ground-truth SH coefficients and the predicted SH coefficients.

For the temporal domain, the HRTF onset features are mostly affected by the listeners' macro-scale dimensions such as head height/width and shoulder width~\cite{xie2013head}. Therefore, the network for onset prediction is designed similarly to the magnitude prediction one in \cref{fig:NN_diagram}. The only difference is on the input side, where it excludes the ear mesh SCH coefficients and the frequency condition, only taking the head and torso's measurements ($x_1$ through $x_{17}$ included in the HUTUBS anthropometric measures\cite{brinkmann2019hutubs}) and ear side as input. The output SH coefficients are computed from the onset values in this case. 

We implemente our deep learning method using PyTorch. For training the deep learning network, we adopt the Adam optimizer. We set the batch size to 1024 and train for 1000 epochs for each training-evaluation trial. We select the model with the lowest validation loss for evaluation. In our experimental setup, the HRTF magnitudes are processed using a dB scale to account of the perceptual considerations described in Section~\ref{ssec: compact_glob_hrtf}. Instead of directly predicting HRTF, we predict the SH coefficients $\hat{\boldsymbol{c}}$. Hence, we multiply the SH basis values $\boldsymbol{Y}$ with the coefficients $\boldsymbol{c}$ to compute the predicted HRTF according to Eq.~\eqref{eq: linear_eqn1}.

\subsection{Dataset}
We use the HUTUBS database~\cite{brinkmann2019hutubs} in our experiments, which includes both the HRTF data and the subjects' scanned head mesh data. We perform experiments and evaluations using the 53 subjects with valid hand scans out of a total of 96 subjects. Part of the anthropometrics data ($x_1$ through $x_{17}$ in the HUTUBS anthropometric measures) included in the database are also used. Since the number of subjects is small, the leave-one-out cross-validation is adopted in this study.

\subsection{Evaluation metrics}
We adopt the log-spectral distortion (LSD) metric to evaluate the performance of our method. The evaluation of the prediction performance is based on these corrected frequency bins mentioned in Section~\ref{ssec:mesh_para}. The LSD can be formulated as:
\begin{equation}
\textit{LSD}(\mathrm{H}, \hat{\mathrm{H}})=\sqrt{\frac{1}{SK} \sum_{s}\sum_{k}\left(20 \log _{10} \left| \frac{H(s, k)}{\hat{H}(s, k)} \right| \right)^{2}},
\label{eq: lsd}
\end{equation}
where $s$ indicates the spatial location, $k$ indicates the frequency index. $S$ and $K$ are the numbers of spatial locations and frequencies, respectively. $H(s, k)$ and $\hat{H}(s, k)$ denote the magnitude of the ground-truth HRTF and the predicted HRTF, respectively, in the linear scale. With HUTUBS setup, the $S$ spatial locations cover the entire discretized space, hence the LSD evaluates global performance.

\begin{figure}[htbp]
    \centering
    \begin{subfigure}[b]{\columnwidth}
        \includegraphics[width=\textwidth]{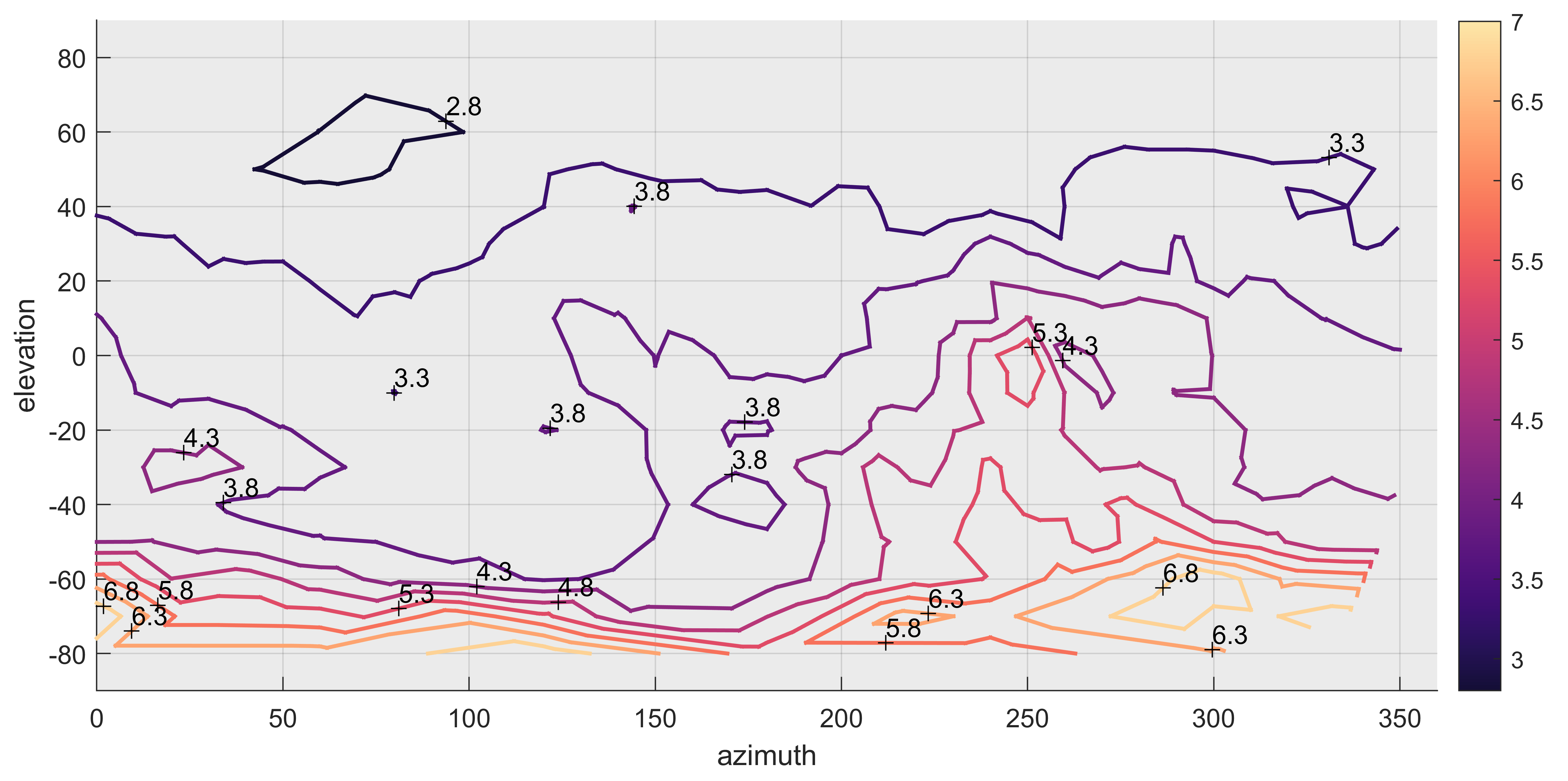}
        \caption{}
    \end{subfigure}

    \begin{subfigure}[b]{\columnwidth}
        \includegraphics[width=\textwidth]{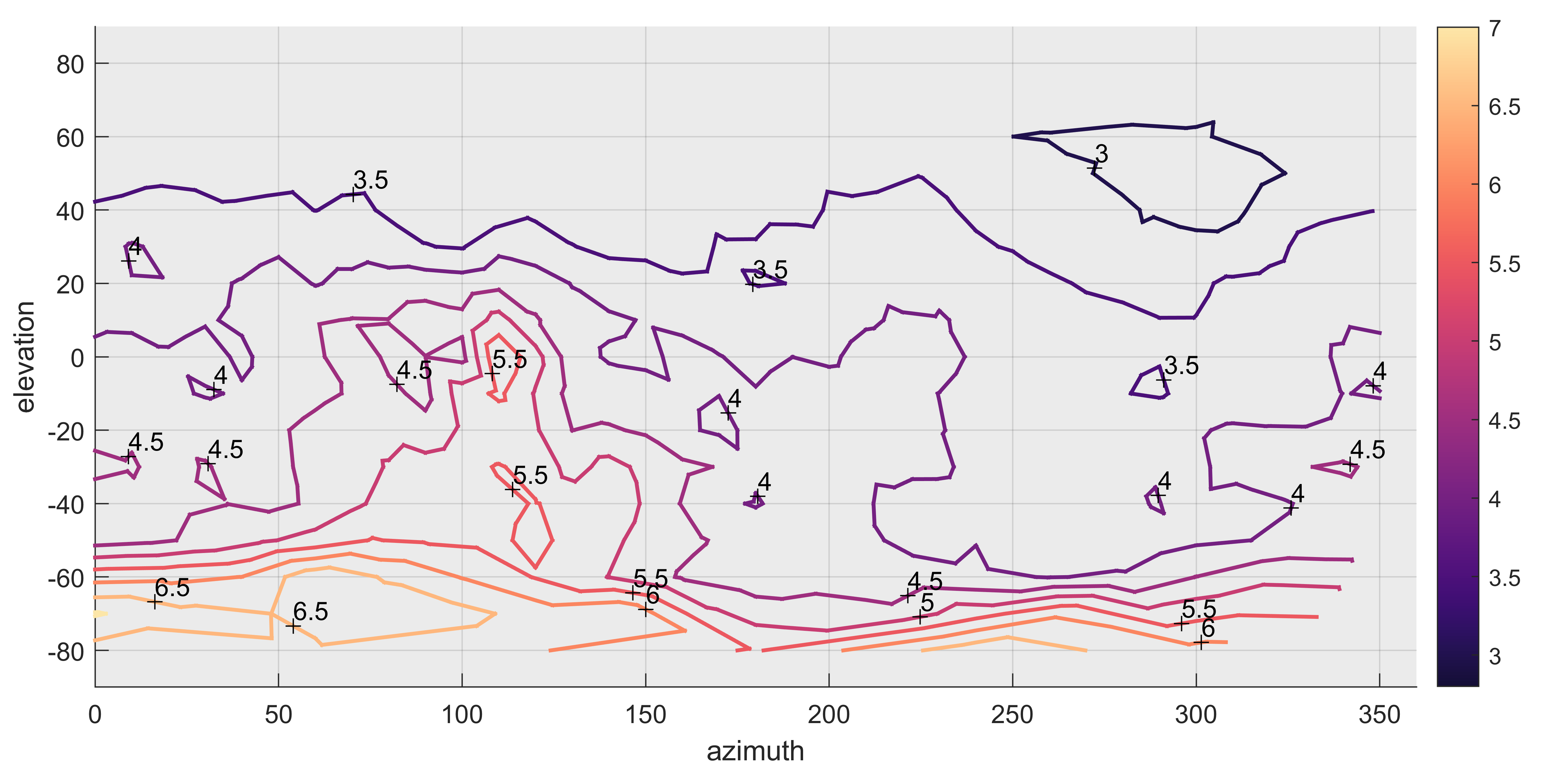}
        \caption{}
    \end{subfigure}
    
\subfigsCaption{Contour plot for global distribution of the LSD between our model's prediction and the original measured HRTFs across all subjects, all frequencies combined. (a) is the left ear case, and (b) is the right ear case.}
\label{fig:LSD_global}
\end{figure}

\begin{figure}[htbp]
\centering
\includegraphics[width=\columnwidth]{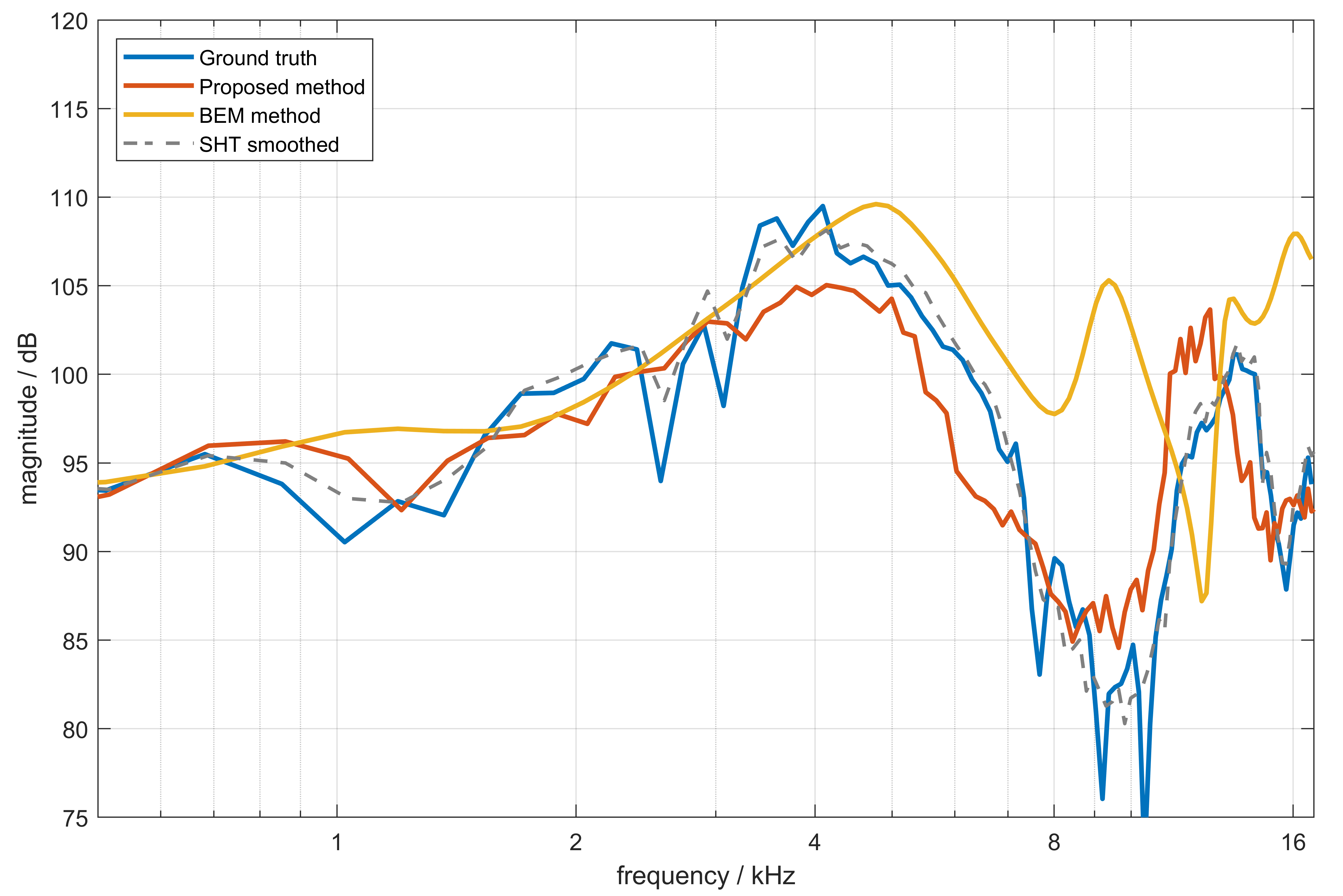}
\caption{Comparison of HRTF prediction from our proposed method, BEM method, to the ground-truth HRTF magnitude spectra at the frontal direction for HUTUBS subject 69. The SHT pre-processed result is also included as the dashed line.}
\label{fig:frontal_comparison_3}
\end{figure}

\section{Experiments and Results}
\label{sec:pred_result}

\subsection{HRTF magnitude prediction results}
The global prediction error distribution is shown in \cref{fig:LSD_global}, compared to the original measured HRTF without processing. Across all the subjects, we achieve mean and standard deviation of ($3.92 \pm 0.83$) dB global LSD for left ears, and ($4.06 \pm 0.78$) dB for right ears, across all source locations and frequencies combined. These results outperform some previous method that uses anthropometric data only, which produced a global LSD of 4.74dB~\cite{wang2020global}. In certain areas on the ipsilateral side (around +90 degrees in azimuth for the left ear, around +270 degrees for the right ear), the LSD reaches the sub-3dB level. For the ipsilateral side, the HRTF magnitudes are generally higher, and the directivity patterns have a distinct main lobe (shown in \cref{fig:HRTF_SHT_4fig} (a) in brighter color area), which are relatively easy for the SHT coefficients to follow, as well as for the algorithm to learn. The errors peak at low elevation regions, especially for the contralateral side (around +270 degrees in azimuth for the left ear, around +90 degrees for the right ear). This might be due to the relatively complex directional patterns in the area where the magnitudes are low (shown in \cref{fig:HRTF_SHT_4fig} (a) in darker color areas), which adds to the difficulty for the prediction algorithm to learn, especially after the smoothing from the SHT processing.

\subsection{HRTF onset prediction results}
Combining all subjects, the onset prediction error is (45.74, 41.57) $\mu s$ for the left/right ear, compared to (13.84, 13.71) $\mu s$ for the error brought by the SHT pre-processing itself. This shows the onset prediction has a decent performance compared to the ground truth across all spatial locations. 

We calculate the LSD for our predicted HRTF and the simulated HRTF using the acoustic boundary element method (BEM) provided by the HUTUBS database, both compared to the ground-truth HRTF. Averaged across all subjects, the LSD is 3.97 dB for our prediction in the frontal direction, and is 6.51 dB for the BEM method. Note that both the simulation and our prediction use the same scanned mesh as the calculation basis, which implies that our method is free from the limitations of the simulation method (e.g., dependency on the solver equations, number of elements, and boundary parameters\cite{xie2013head}) and produces better results at the frontal direction. An example of comparing the HRTF predictions from our proposed method and BEM method to the ground truth at frontal location, where $s = (\theta, \varphi) = (0,0)$, is demonstrated in Figure \ref{fig:frontal_comparison_3}. The SHT smoothed HRTF is also shown as the dashed grey line. The SHT pre-processing produces a smoothed version of the ground truth HRTF, and the prediction HRTF follows the overall trend well, while deviations can be observed at higher frequencies. The BEM result performs well at low frequencies, but has major errors at high frequencies.

\begin{figure}[htbp]
\centering
\includegraphics[width=\columnwidth]{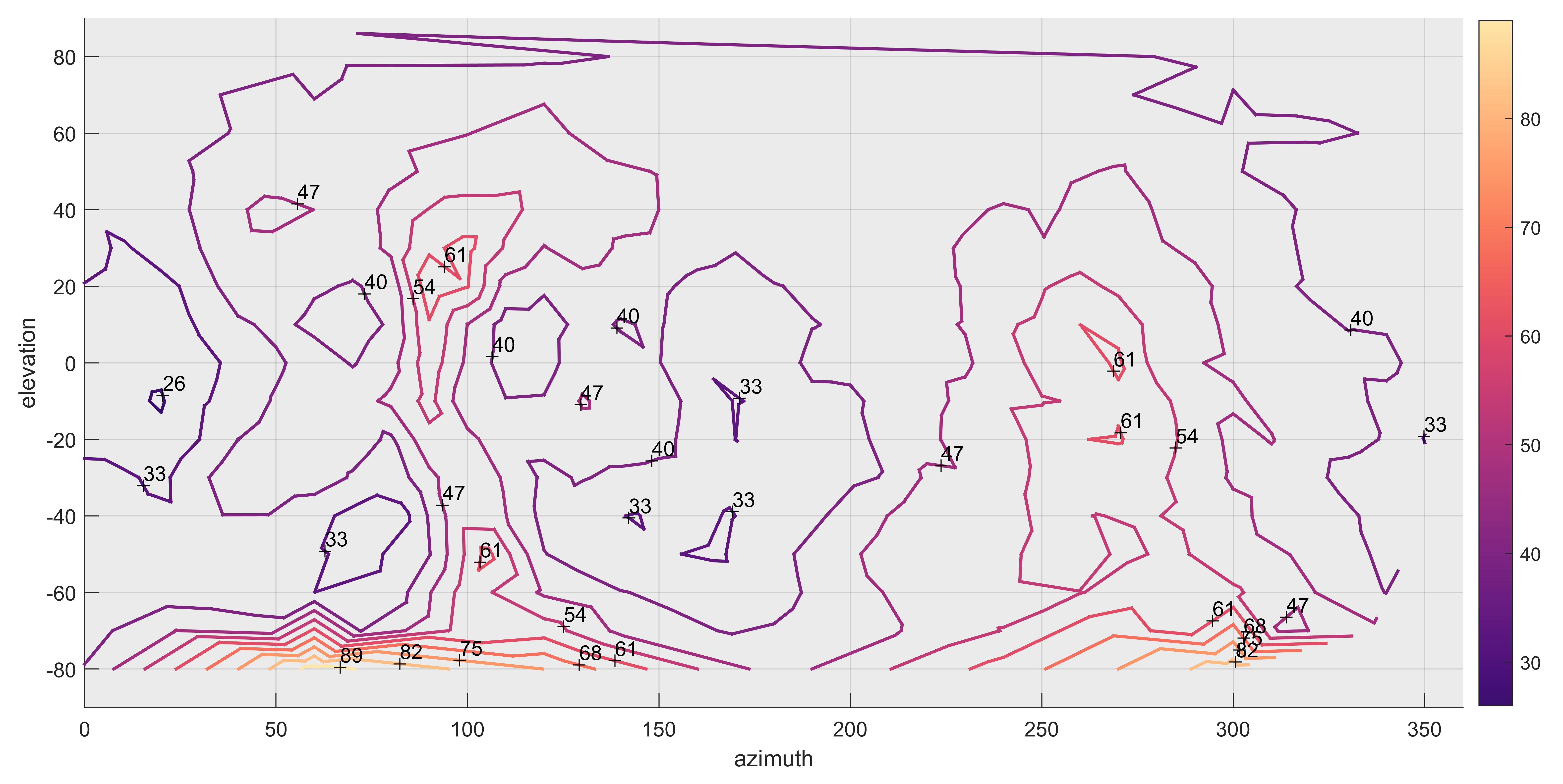}
\caption{Contour plot for global distribution of the ITD prediction error in $\mu s$ across all subjects.}
\label{fig:ITD_error_global}
\end{figure}

We continued to evaluate ITD results by subtracting left onsets from right ones according to the definition. The prediction achieves a mean and standard deviation of $(44.18 \pm 8.20)$ $\mu s$ for ITD error averaged across all locations and all subjects. Although ITD is not the learning target in the neural network design, the ITD predictions closely match the original ones for various elevation cases. This result may have benefited from the efficient feature extraction of the global onset patterns. In \cref{fig:ITD_error_global}, we show the global prediction error of ITD in $\mu s$ at different azimuth and elevation angles. The predicted ITDs generally have smaller errors than the average just-noticeable difference (JND) for human subjects perceiving ITDs across all spatial locations~\cite{andreopoulou2017identification}, especially at frontal areas (JND is about 40 $\mu s$) and lateral areas (JND is more than 100 $\mu s$). The prediction performance in the medial regions (area around 0 and +180 degrees in azimuth) is generally better than that in the lateral regions (area around +90 and +270 degrees in azimuth). This could be because both the onset/ITD peak values occur at lateral regions, where the pre-processing may produce larger smoothing errors. Lower elevation regions (below -70 degree elevation) also have larger errors, and it might be due to the similar reason for the magnitude case that the values in that area have more noisy patterns, as the lower magnitudes cause unstable onset detection.

\begin{figure}[htbp]
    \centering
    \begin{subfigure}[b]{\columnwidth}
        \includegraphics[width=\textwidth]{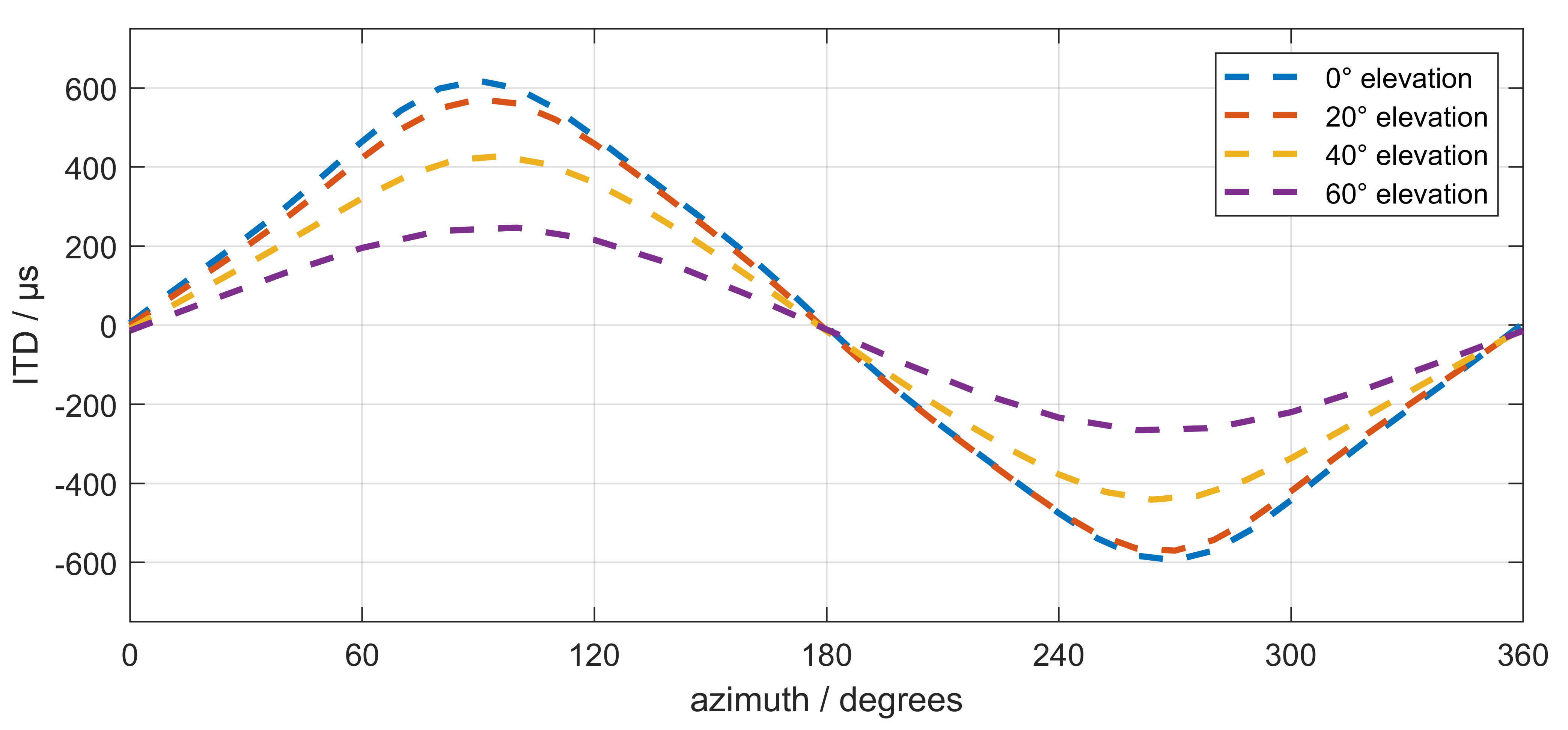}
        \caption{}
    \end{subfigure}

    \begin{subfigure}[b]{\columnwidth}
        \includegraphics[width=\textwidth]{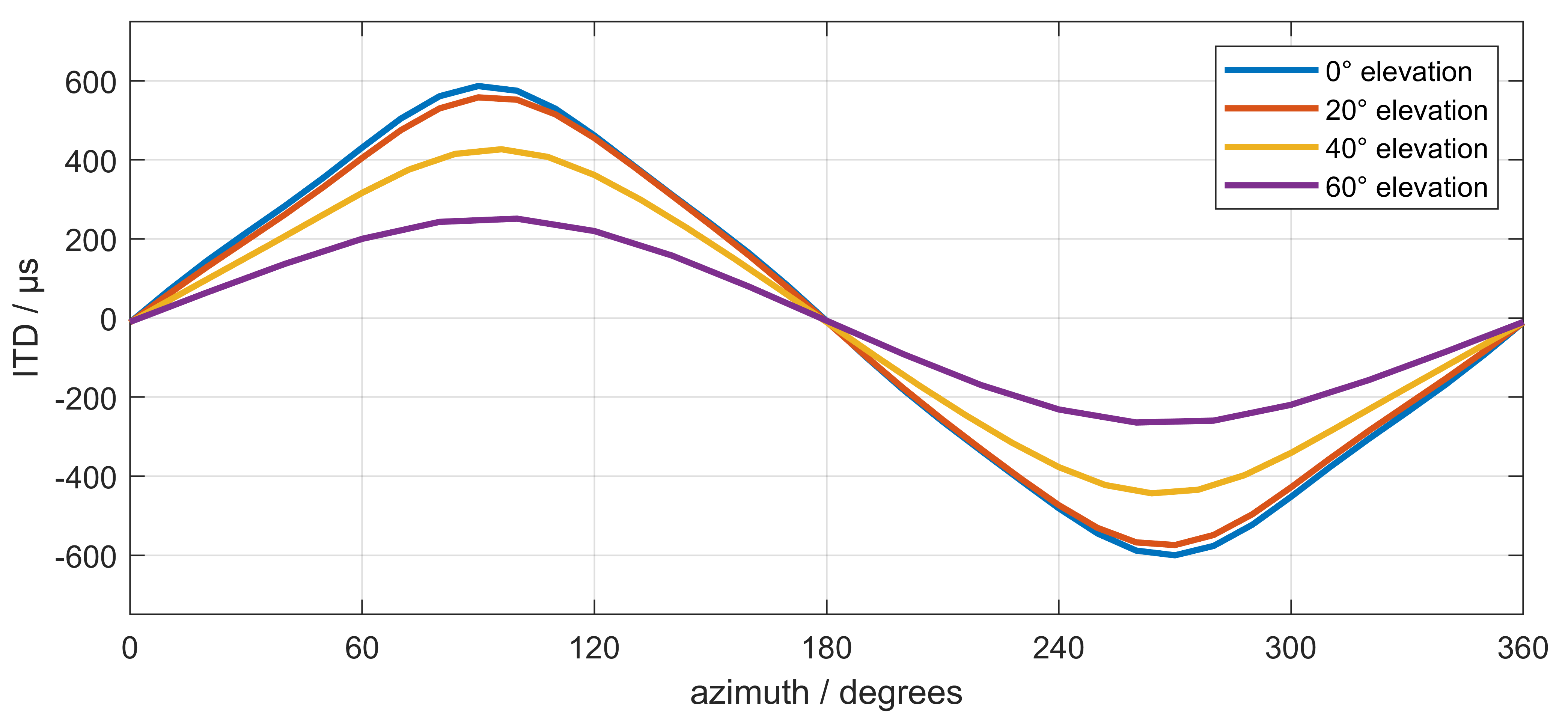}
        \caption{}
    \end{subfigure}
 
\subfigsCaption{ITD prediction result. (a) is one subject's original ITDs at {0, 20, 40, 60} degrees elevations, and (b) is the corresponding predicted ITDs.} 
\label{fig:itd_pred_demo}
\end{figure}

In \cref{fig:itd_pred_demo} we show an example of a comparison of the original (dashed lines) and predicted ITDs (solid lines) of one subject at multiple elevation angles. Comparing the prediction to the original case, the ITD trends and values match well across the azimuth directions in all elevation cases.

\begin{figure*}[t!]
    \centering

    
\subfloat[]{\includegraphics[width=0.5\columnwidth]{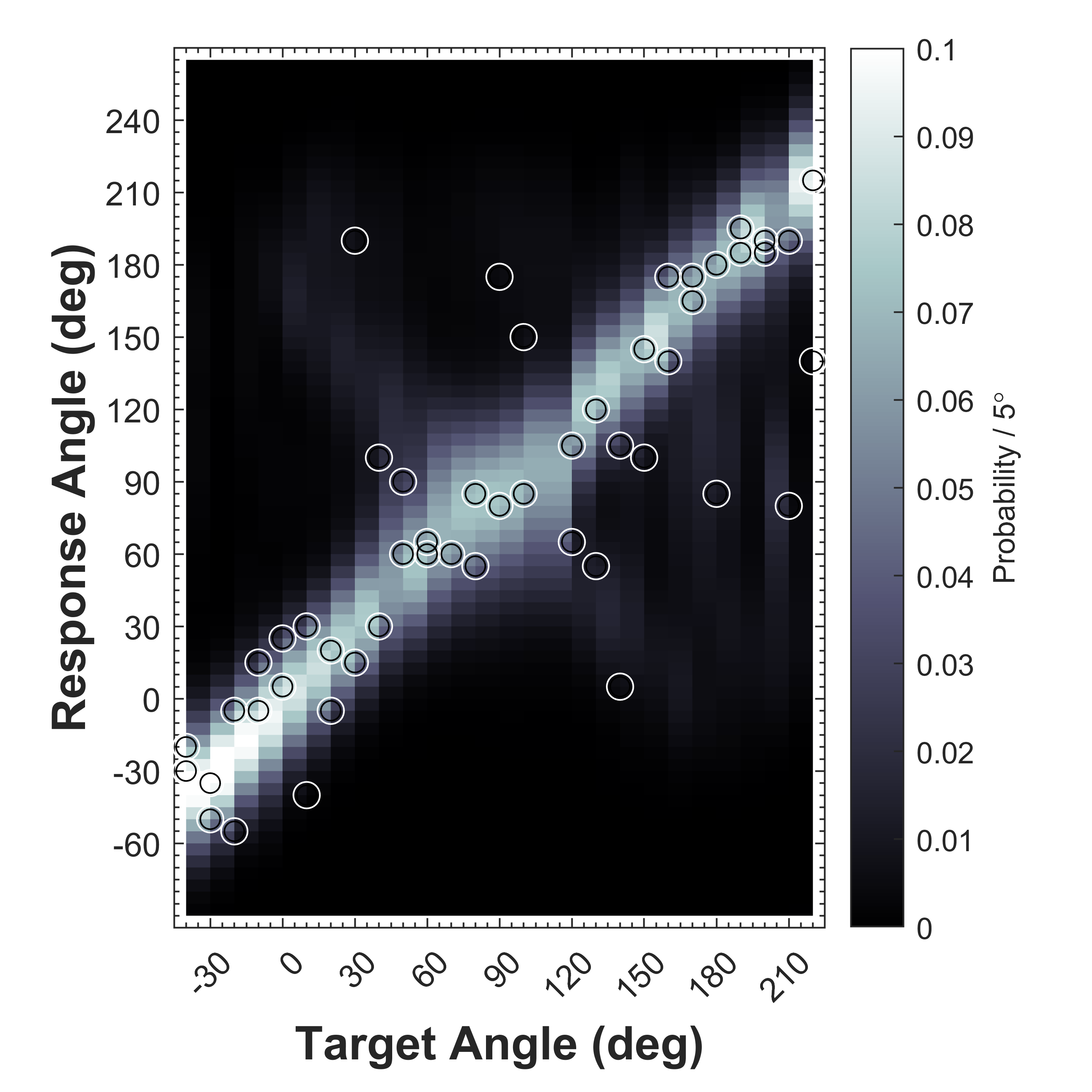}%
\label{baumgartner_first_case}}
\subfloat[]{\includegraphics[width=0.5\columnwidth]{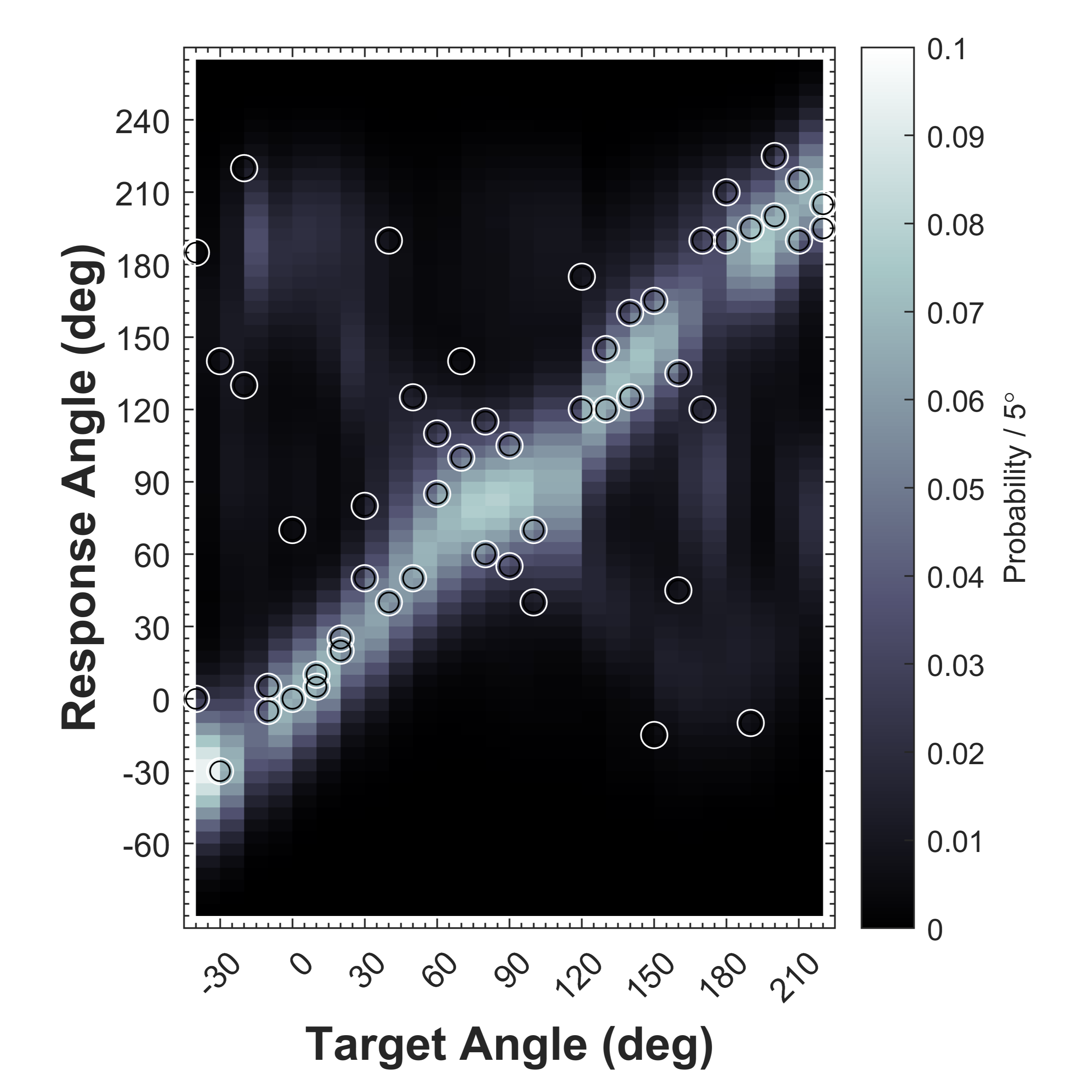} %
\label{baumgartner_second_case}}
\subfloat[]{\includegraphics[width=0.5\columnwidth]{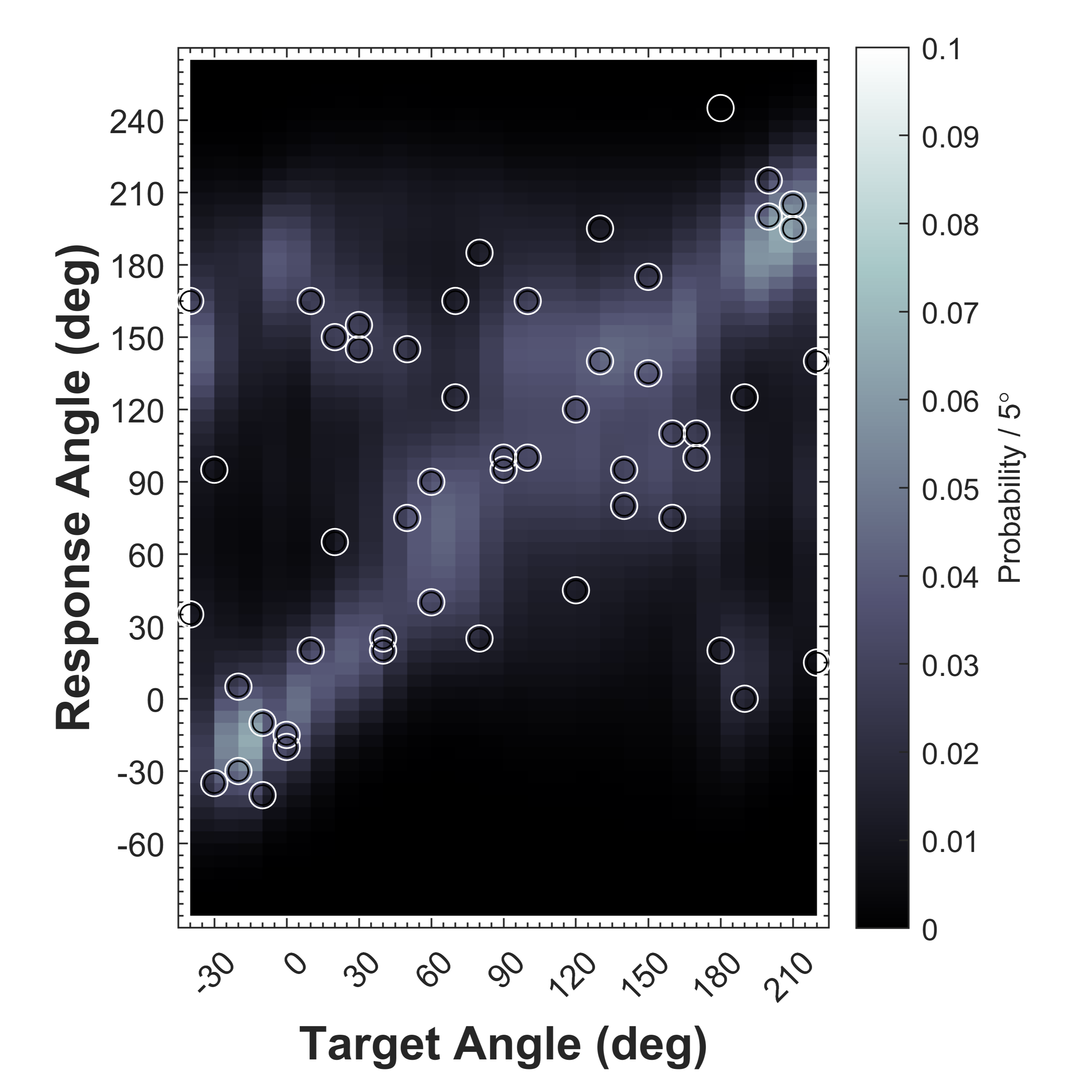} %
\label{baumgartner_3rd_case}}
\subfloat[]{\includegraphics[width=0.5\columnwidth]{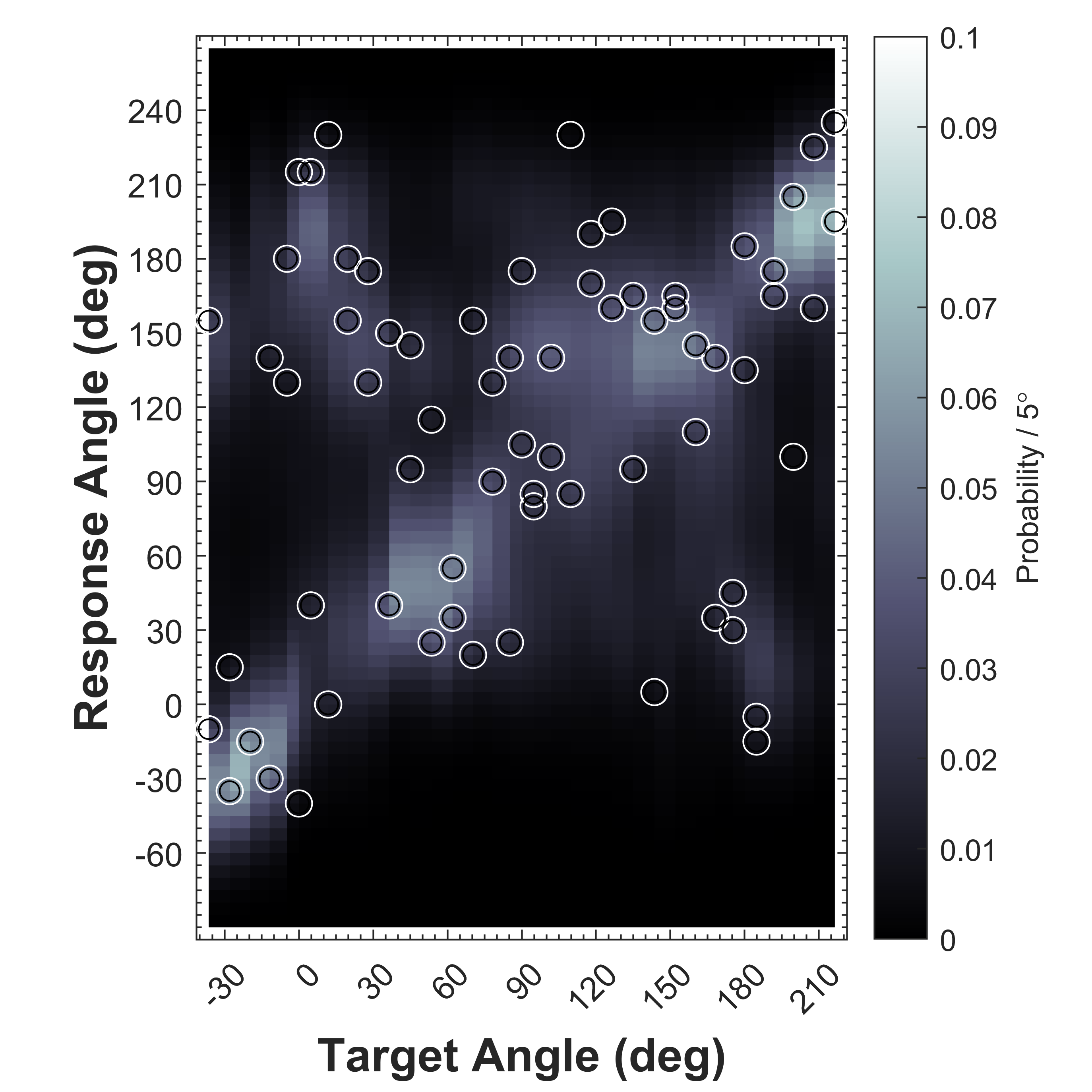} %
\label{baumgartner_4th_case}}
\subfigsCaption{Localization simulation results comparison using the sagittal-plane localization model~\cite{baumgartner2014modeling}. The response angle (y axis) is shown as a function of the target angles (x axis), and the probability is coded as brightness. The circles represent the estimated localized positions for the listener. (a) is the result with the original measured HRTF, (b) is the smoothed, (c) is our prediction, and (d) is the BEM simulated provided by HUTUBS.}
\label{fig:baumgartner2014_model_demo}
\end{figure*}

\subsection{Localization simulation results}
We performed perceptual experiment simulations using a localization model implemented in the auditory modeling toolbox (AMT)~\cite{Sndergaard2013}. This model has been widely adopted to estimate the probability of user responses across sagittal planes~\cite{baumgartner2014modeling, Miccini2021, Pollack2021}. We used the lateral offset of 0 in the setup. Across all subjects, the localization performance was evaluated by means of the quadrant error rates (QEs) and polar errors (PEs)~\cite{baumgartner2014modeling}. The QE describes 
the rate of the simulated localization errors greater than 90 degrees. The PE describes the root-mean-square error of simulated responses within the correct hemiﬁeld. 
The averaged QE (in percentage) and PE (in degrees) values are (11.62, 29.22) for original measured HRTFs, (17.79, 32.22) for smoothed ones, (36.49, 44.94) for our predicted ones, and (41.93, 51.12) for the BEM simulated ones. 

The comparison of the localization result for a subject's original measured, smoothed, predicted, and BEM simulated HRTFs are shown in \cref{fig:baumgartner2014_model_demo}. From the figure, it can be seen that the original HRTF in (a) has good localization performance. The smoothed case in (b) is slightly worse, due to the smoothing effect from the SHT processing. The estimated response angles (the brightness and circles) in (c) are more centralized around the target angles, compared to the BEM case in (d). These findings indicate that although the result of the predicted HRTFs is worse than the smoothed ones, it is still better than those of the BEM simulation. There are a considerable amount of front-back confusion for both the predicted case and the BEM case in this localization experiment simulation. These results are consistent with the QE and PE values reported above for each case.

\section{Conclusions and Discussions}
\label{sec:discussion}
In this paper, we presented our novel approach to predicting global HRTFs from the subject's scanned ear mesh and head geometry. 
We explored efficient methods to represent both HRTFs and scanned mesh in compact forms, and designed the machine learning framework to find the connection between intrinsic features of acoustics and geometry dimensions. 
For the HRTF pre-processing step, we extract the key information in HRTF datasets from frequency and temporal domain. For the mesh processing step, the SCHA method sufficiently preserved features within the ear geometry while using a substantially small amount of data. Benefiting from the efficient data representation and network design, our proposed method produces better prediction results in both frequency and temporal domains across spatial locations compared to some previous work which only included anthropometric measurements as prediction input. 

With the learning target to be the SHT coefficients rather than HRTF values directly, the algorithm is able to capture the overall pattern with high efficiency, in both time and frequency domains, as SHT representations only require relatively low truncation order to recover the global HRTF pattern while maintaining perceptual viability. The compact representations also help reduce the computational complexity of the neural network. 

Our proposed method outperforms the anthropometric-data prediction, although it is only trained on 53 individuals while the latter~\cite{wang2020global} is trained on 93 individuals. This suggests that the more detailed geometry information provided by the mesh data helps with the HRTF prediction. In addition, unlike the anthropometric measurements that are manually selected parameters, taking in the mesh information avoids the potential bias in the choice of parameters. 

Only a few existing works employ direct mesh processing within the whole ear region for the neural network in HRTF prediction. A previous work~\cite{Zhou2021} utilized ear mesh voxelization to predict the simulated HRTF spectra, showing the prediction advantages of neural networks in speed and complexity. However, this method neglects the information for the torso and head. The voxelization also required a significant amount of parameters to represent ear geometry. Furthermore, the training dataset was generated by the acoustic simulation method (FDTD), with a much larger size compared to our work. In other previous work~\cite{Fantini2021, Miccini2020}, although the ear mesh was employed as input, the mesh information was used to extract lower-dimension features such as anthropometric measurements, without leveraging the more comprehensive representation offered by 3D geometry. Another work proposed a parametric pinna model to align to new listers pinna and calculate personalised HRTFs~\cite{Pollack2021}, yet resulting in higher geometric error than our method. 

For future work, some subjective perceptual examination of our prediction is desired to show the method's effectiveness in auditory perception. The adoption of a more recent deep learning framework remains to be further explored, which might enable us to perform the prediction from the limited number of subjects more effectively. Also, some insights from the geometric information can be employed in our predictions, considering the acoustic scattering law that connects the scattered energy with the equivalent size of the object. We could employ physics-informed constraints in our neural network to assign weights from geometric features to acoustic features. We would also expand our processing methods to various HRTF databases using appropriate harmonization methods~\cite{Zhang2023}, to alleviate the inconvenience resulting from different measurement standards.



\bibliographystyle{abbrv-doi}

\bibliography{reference}

\end{document}